\documentclass[10pt,doublecolumn,twoside]{IEEEtran}

\usepackage{etoolbox}
\newtoggle{doublecolumn}
\newtoggle{calculateFigures}

\toggletrue{doublecolumn}
\togglefalse{calculateFigures}
\usepackage{etex}

\usepackage[utf8]{inputenc}
\usepackage[T1]{fontenc}

\usepackage[english]{babel}
\usepackage{lipsum}
\usepackage{amsbsy,amsmath,amsfonts,amssymb,amsthm}
\usepackage{mathtools}
\usepackage{textcomp} 
\usepackage{relsize}

\usepackage{bm,cite,cases,pstricks,times,url,verbatim} 
\usepackage[noend]{algpseudocode}
\usepackage{booktabs}
\usepackage{tabularx,array,dcolumn,multirow}

\usepackage{algorithm}

\usepackage{soul} 

\usepackage{blkarray,bigdelim} 

\allowdisplaybreaks[4]

\usepackage{centernot} 

\newcommand{\subparagraph}{}

\iftoggle{doublecolumn}{
    \newtheorem{thm}{Theorem}
    \newtheorem{fact}{Fact}
    \newtheorem{lemma}{Lemma}
    \newtheorem{definition}{Definition}
    \newtheorem{conj}{Conjecture}
    \newtheorem{propos}{Proposition}
    \newtheorem{corol}{Corollary}
    \newtheorem{ass}{Assumption}
    \newtheorem{example}{Example}
    \newtheorem{remark}{Remark}
    \newtheorem{note}{Note}
    \newtheorem{obs}{Observation}

}{

    \newtheoremstyle{exampstyle}
      {0} 
      {0} 
      {\itshape} 
      {} 
      {\bfseries} 
      {.} 
      {.5em} 
      {} 

    \theoremstyle{exampstyle} \newtheorem{thm}{Theorem}
    \theoremstyle{exampstyle} 
    \theoremstyle{exampstyle} \newtheorem{lemma}{Lemma}
    \theoremstyle{exampstyle} 
    \theoremstyle{exampstyle} 
    \theoremstyle{exampstyle} 
    \theoremstyle{exampstyle} \newtheorem{corol}{Corollary}
    \theoremstyle{exampstyle} 
    \theoremstyle{exampstyle} 
    \theoremstyle{exampstyle} 
    \theoremstyle{exampstyle} 
    \theoremstyle{exampstyle} 
}

\makeatletter
\newcommand{\pushright}[1]{\ifmeasuring@#1\else\omit\hfill$\displaystyle#1$\fi\ignorespaces}
\newcommand{\pushleft}[1]{\ifmeasuring@#1\else\omit$\displaystyle#1$\hfill\fi\ignorespaces}

\begingroup
\catcode`\#=11
\gdef\noautorotate{-dAutoRotatePages#/None}
\endgroup

\usepackage{graphicx,xcolor,float,dblfloatfix}
\usepackage{psfrag}

\usepackage{caption}
\usepackage{subcaption}

\iftoggle{calculateFigures}{
    \usepackage[crop=off,runs=2,pspdf=\noautorotate]{auto-pst-pdf}
}{
    \usepackage[off]{auto-pst-pdf}
}

\newcommand{\subalign}[1]{%
  \vcenter{%
    \Let@ \restore@math@cr \default@tag
    \baselineskip\fontdimen10 \scriptfont\tw@
    \advance\baselineskip\fontdimen12 \scriptfont\tw@
    \lineskip\thr@@\fontdimen8 \scriptfont\thr@@
    \lineskiplimit\lineskip
    \ialign{\hfil$\m@th\textstyle##$&$\m@th\textstyle{}##$\crcr
      #1\crcr
    }%
  }
}

\usepackage{hyperref} 

\graphicspath{%
{./Figures/}
}

\newcommand\ddfrac[2]{\frac{\displaystyle #1}{\displaystyle #2}} 

\usepackage{color}
\newcommand{\AB}[1]{{\color{red} \ul{\textbf{[AB Comment]:}} #1}}

\newcommand{\tth}{{\rm th}}
\newcommand{\een}{{\rm en}}
\newcommand{\ttx}{{\rm tx}}

\usepackage[compact]{titlesec} 

\begin{document}

\author{\IEEEauthorblockN{Alessandro~Biason,~\IEEEmembership{Student~Member,~IEEE,} Chiara~Pielli, \\Andrea~Zanella,~\IEEEmembership{Senior~Member,~IEEE,}  and~Michele~Zorzi,~\IEEEmembership{Fellow,~IEEE}
\thanks{The authors are with the Department of Information Engineering, University of
Padua, 35131 Padua, Italy (email: biasonal@dei.unipd.it; piellich@dei.unipd.it; zanella@dei.unipd.it; zorzi@dei.unipd.it).}
\thanks{A preliminary version of this paper has been accepted for publication at IEEE GLOBECOM 2016~\cite{Pielli2016}.}
}}

\title{On the Energy/Distortion Tradeoff in the IoT}

\maketitle
\pagestyle{empty}
\thispagestyle{empty}


\begin{abstract}
The Internet of Things paradigm envisages the presence of many battery-powered sensors and this entails the design of energy-aware protocols. Source coding techniques allow to save some energy by compressing the packets sent over the network, but at the cost of a poorer accuracy in the representation of the data.
This paper addresses the problem of designing efficient policies to jointly perform processing and transmission tasks. In particular, we aim at defining an optimal scheduling strategy with the twofold ultimate goal of extending the network lifetime and guaranteeing a low overall distortion of the transmitted data. We propose a Time Division Multiple Access (TDMA)-based access scheme that optimally allocates resources to heterogeneous nodes. We use realistic rate-distortion curves to quantify the impact of compression on the data quality and propose a complete energy model that includes the energy spent for processing and transmitting the data, as well as the circuitry costs. 
Both full knowledge and statistical knowledge of the wireless channels are considered, and optimal policies are derived for both cases.
The overall problem is structured in a modular fashion and solved through convex and alternate programming techniques. Finally, we thoroughly evaluate the proposed algorithms and the influence of the design variables on the system performance adopting parameters of real sensors.
\end{abstract}

\section{Introduction}\label{sec:introduction}

The Internet of Things (IoT) is a recent communication paradigm that realizes the concept of ubiquitous computing and has found application in a large number of fields, e.g., environmental monitoring, smart buildings, and health care~\cite{miorandi}. 
In many IoT scenarios, devices are likely going to be battery-powered and thus rely on a limited energy supply. It is of paramount importance to develop protocols and algorithms that are energy-centric, as replacing batteries when depleted may be a costly and difficult operation~\cite{lazarou}.

In monitoring and sensing applications, the devices constantly sense the surrounding environment and exchange a large amount of raw data, whose transmission would rapidly deplete the batteries of the nodes. A way to save some transmission energy consists in reducing the volume of data to send by applying source processing techniques. Typically, the signals generated by the nodes exhibit time correlation, thus the use of lossy compression may be very helpful to eliminate the intrinsic data redundancy. However, this comes at the cost of spending some energy for the compression operations and introducing a distortion with respect to the original signal.
In this work, we address the problem of scheduling transmissions for a heterogeneous IoT network 
with special focus on the energy dynamics and the quality of the transmitted information, with the goal of determining the optimal tradeoff between these two aspects.

The need for energy efficiency in IoT and Wireless Sensors Networks (WSNs) has been gaining increasing attention in the last years, and a large variety of energy-aware protocols at all layers of the protocol stack have been proposed. To extend the lifetime of the network as much as possible, two major directions exist: when the battery represents the only source available to the device, the effort is put in the minimization of the energy consumption~\cite{anastasi,kouyoumdjieva}; otherwise, when devices have Energy Harvesting (EH) capabilities, energy-neutral operation modes are sought~\cite{gunduz,ulukus2015,Biason2015d}. In this work, we follow the first branch and pursue minimization of the energy consumption under Quality of Service (QoS) requirements.
Many works in literature study the intrinsic tradeoff between energy efficiency and QoS in WSNs. In~\cite{chen2010}, an adaptive fault-tolerant QoS control algorithm has been developed with the goal of prolonging the lifetime of query-based sensor networks as much as possible. Often, QoS metrics drive the design of energy-efficient routing algorithms that optimize the path survivability of WSNs, and that typically leverage on the spatial correlation between the sensors' data or the asymmetry in capabilities and resources available to the devices~\cite{pantazis, jabbar}. The benefits of network topology control on the energy minimization have also been the object of research analysis~\cite{Deshpande, Wightman}. Other works investigate the tradeoff between energy consumption and delay in the data delivery, often exploiting temporal correlation of the data~\cite{Fateh}, or using transmission range slicing~\cite{Ammari}.
A comprehensive survey on strategies to prolong the lifetime of WSNs is provided in~\cite{Curry}.

We focus on the design of the Medium Access Control (MAC) layer, which has a strong impact on the energy efficiency, since the usage of the RF chain may be very energy-demanding~\cite{alameen}.
Although random access is sometimes preferred over coordinated access~\cite{michelusi, zanella2013}, TDMA-based schemes have proved to be a valid choice in IoT contexts~\cite{berger2012,lee}. Contention-based protocols are flexible and require low synchronization costs, but generally lead to a high energy wastage due to collisions and idle listening, which can be instead avoided in reservation-based protocols, at the cost of some additional synchronization overhead~\cite{bachir}. In applications like environmental monitoring, the set of nodes involved in the data reporting operation is usually fixed, and sensors typically report data periodically. Thus, the traffic pattern is known in advance, which makes TDMA well suited for this reporting regime.  
Indeed, by properly matching the slots allocation to the traffic pattern, idle listening and collisions can be completely avoided, thereby prolonging the lifetime of the network. To this aim, appropriate duty cycling mechanisms are typically adopted.
In~\cite{wu}, the authors propose a TDMA scheme where nodes are scheduled with consecutive time slots in different radio states, thereby reducing the energy costs due to synchronization. 
The SAS-TDMA protocol~\cite{Shen2013} is a source-aware scheduling algorithm that adapts itself to the network dynamics to improve the QoS in real-time industrial wireless sensor and actuator networks.
Often, TDMA is combined with Channel Sensing Multiple Access (CSMA) techniques, since these hybrid approaches offer flexibility when choosing the frame length and assigning slots to nodes~\cite{jovanovic,lenka}. A comprehensive survey of MAC protocols for WSNs is given in~\cite{huang2013}. 

The example that most corroborates the validity of TDMA in IoT networks is the Time-Slotted Channel Hopping (TSCH) mode of the IEEE~802.15.4e standard~\cite{tsch}. It was introduced in 2012 by the Internet Engineering Task Force, and it currently represents the major emerging standard for industrial automation and process control Low-power and Lossy Networks (LLNs). TSCH uses time synchronization to achieve ultra low-power operation and channel hopping to enable high reliability by mitigating the effect of narrow-band interference and multi-path fading.

The design of our MAC scheme revolves around the performance metric of distortion. Other works in the literature consider joint source coding and transmission policies and study the tradeoff between energy efficiency and data quality~\cite{chidean, castiglione}.
In~\cite{tapparello}, an online joint compression and transmission optimization strategy is investigated for sensors with EH capabilities that generate correlated information, but how to schedule transmissions in a time slot is not treated. In~\cite{zordan_comp}, the authors derive optimal compression policies for a single sensor in order to minimize the long-term average distortion subject to the energy sustainability of the sensor, where power control is used to adapt the transmission to the status of the fading channel. In~\cite{dey}, energy allocation strategies are proposed with the goal of minimizing the signal distortion when several sensors measure the same process of interest and exploit data fusion techniques, but analytical results are derived only for a two-node system. Finally,~\cite{bing} proposes a TDMA scheduling where times slots are allocated in a dynamic fashion based on the spatial correlation of the transmitted signals. 

In this work, we aim at determining the optimal operating point in the tradeoff between network lifetime and signal quality in order to derive a TDMA-based transmission scheduling strategy of the resource-constrained nodes. 
With respect to our previous work~\cite{Pielli2016}, we introduce two major novelties. First, we allow for the dismission of some users from transmission when it is impossible to have all devices meet their requirements in a frame. Then, we relax the full CSI assumption, since in realistic scenarios with fast fading it is impractical to perfectly know the channel realizations a priori.  Hence, we consider the case in which devices have only statistical CSI for future slots. 

The major contributions of our work are summarized as follows.
\begin{itemize}
 \item We set up a scheduling problem in which data compression and transmission are jointly optimized, in a scenario with multiple sensors that communicate to the same gathering point. We account for realistic rate-distortion curves that accurately match those of practical data compression algorithms. Futher, we introduce an exhaustive parametric model of the energy dynamics of a device, considering all the main sources of energy consumption;
 
 \item We structure the overall complex problem into interrelated subproblems. This modular structure makes the framework more robust and flexible, as the different parts can be modified in an independent fashion, provided that the relation between the blocks is maintained;
 
 \item We derive a set of theoretical results that allow us to define optimal policies when CSI is known perfectly (full knowledge) or statistically;
 
 \item We conduct a thorough numerical evaluation of the proposed policies, considering realistic parameters and describing the effect of the various system variables on the proposed framework, gauging their influence on the performance metrics. Moreover, we emphasize the importance of using the optimal policy over other simpler schemes. 
\end{itemize}

The rest of the paper is organized as follows. In Section~\ref{sec:system_model} we describe our system model and introduce the optimization problem, which is made up of two parts, namely the Frame-Oriented Problem and the Energy-Allocation Problem. Sections~\ref{sec:FOP} and~\ref{sec:FOP_avg_CSI} solve the Frame-Oriented Problem when fast fading is neglected or taken into account, respectively, and some common aspects of the two versions of the problem are discussed in Section~\ref{sec:FOP_notes}, which also introduces the user dismission policy. The Energy-Allocation Problem is studied in Section~\ref{sec:EAP}. In Section~\ref{sec:numerical_evaluation} we show the numerical results of the proposed policies, and, finally, Section~\ref{sec:conclusions} discusses the proposed solutions and gives an overview of the possible extensions of the work.

\emph{Notation:} Matrices and vectors are represented with boldface letters, and the subscript and superscript refer to the row and column index, respectively; accordingly, $\mathbf{E}_i$ refers to the $i$-th row of matrix $\mathbf{E}$, $\mathbf{E}^{(k)}$ to its $k$-th column, and $E_i^{(k)}$ is the $(i,k)$ element. With ``$\forall i$'' and ``$\forall k$'', we summarize $i \in \mathcal{N} \triangleq \{1,\dots,N\}$ and $k \in \{1,\ldots,n\}$, respectively, where $N$ and $n$ are defined in the next section.

\section{System model}\label{sec:system_model}
We consider $N$ heterogeneous sources that wirelessly send data to a central Base Station (BS). 
Users access the uplink channel in a TDMA fashion, and time is partitioned into frames, where frame $k$ corresponds to the time interval $[t_k,t_{k+1})$.
Each node periodically generates data, decides whether and how much to compress it, and finally transmits it to the common receiver.

\subsection{Data Generation and Compression}\label{subsec:data_model}
Nodes may generate data by collecting measurements from the surrounding environment or by serving as relays to the common receiver for farther nodes. The size of the data generated in frame $k$ by node $i$ is denoted as $L_{0,i}^{(k)}$.
Each user $i \in \mathcal{N}$ is capable of compressing its data using a lossy compression scheme, which may be source-specific. The compression operation affects the quality of the transmitted information and introduces a distortion $D_i^{(k)}$, which is a function of the compression ratio $\eta_{P,i}^{(k)} =  L_{i}^{(k)}/L_{0,i}^{(k)}$, where  $L_i^{(k)} \le L_{0,i}^{(k)}$ is the size of the compressed packet in frame $k$. It is thus possible to define a function that maps the distortion to the transmission rate or, equivalently, to the compression ratio. 
Typically, closed-form expressions for the rate-distortion curves are only available for idealized compression techniques operating on Gaussian information sources
, whereas for practical algorithms such curves are generally obtained experimentally. Leveraging on the work in~\cite{zordan}, we account for realistic rate-distortion curves that can be mathematically modeled as follows:~
\begin{align}
    D_i^{(k)} =  \left[b_i \left(\dfrac{1}{(\eta_{P,i}^{(k)})\,^{a_i}} - 1\right) \right]^+ ,
    \label{eq:dist}
\end{align}

\noindent where $a_i, b_i > 0$ and $[\cdot]^+ \triangleq \max\{\cdot,0\}$. 
Notice that the distortion is zero when the packet is not compressed, i.e., $\eta_{P,i}^{(k)}=1$.

We assume that the data collected in a frame is lost if not transmitted in the next frame, which is equivalent to imposing a strict limit on latency, or to consider finite data buffers at the devices.

\emph{\textbf{QoS Requirement}.} We also introduce a QoS requirement on the data quality: $D_i^{(k)} \leq D_{\tth, i}^{(k)}$, where $D_{\tth, i}^{(k)}$ is a threshold distortion level. If the reconstruction error exceeds this threshold, the signal generated by the source node is no longer useful for the final destination. Throughout this paper, we always use distortions that satisfy this requirement, and neglect other results.

\subsection{Channel model}

The average physical rate of user $i\in\mathcal{N}$ in frame $k$ is given by Shannon's bound\footnote{Actually, the SNR should be scaled by a proper margin factor to account for the gap between the spectral efficiency of practical modulation schemes and the Shannon bound, but for the sake of simplicity we neglect this constant term, without undermining the generality of the model. 
}:~
\begin{align}
    r_i^{(k)} = W \log_2 \big( 1+\gamma_i^{(k)} \big) = W \log_2 \big( 1+h_i^{(k)}\,P_i^{(k)} \big),   \label{eq:rate}
\end{align}

\noindent where $W$ is the bandwidth, $\gamma_i^{(k)}$ the Signal-to-Noise Ratio (SNR), $P_i^{(k)}$ the transmission power, and $h_i^{(k)}$ the channel gain normalized with respect to noise. 
The $N$ channel gains $h_1^{(k)},\ldots,h_N^{(k)}$ are affected by fast fading, which evolves independently over time and is independent among users.

In Sections~\ref{sec:FOP} and~\ref{sec:FOP_avg_CSI}, we will investigate the compression-transmission problem without and with explicitly accounting for the fading, respectively. Although the first approach produces much simpler policies, it may be strongly suboptimal if applied to the more realistic cases with fading (further details will be given in Section~\ref{subsec:sub_opt_pol}).

\subsection{Energy Consumption Model}\label{subsec:energy_model}

We consider all devices to be battery-powered and denote as $B_i^{(k)}$ the battery level of node $i$ in frame $k$. 
The initial battery level $B_i^{(0)}$ represents the only energy available to node $i$, which therefore has a strong impact on the system performance.
In every frame, a non-negative amount of energy $E_i^{(k)}\in [0, B_i^{(k)}]$ is used for processing and transmission tasks. In the following, we characterize the diverse sources of energy consumption we account for.

\emph{\textbf{Data processing}.}
There exists no universal way to parametrically characterize the energy consumed for compression; to model it, we exploit the results of~\cite{zordan}, and express it as:~ 
 \begin{align}
   E_{P,i}^{(k)}  = E_{0,i} \cdot L_{0,i}^{(k)} \cdot N_{P,i}(\eta_{P,i}^{(k)}),    \label{eq:e_processing}
 \end{align}
 \noindent where $E_{0,i}$ is the energy consumption per CPU cycle and depends on the processor of the device, and function $N_{P,i}(\eta_{P,i}^{(k)})$ represents the number of clock cycles per bit needed to compress the original signal, which depends on the compression ratio. Different compression algorithms entail different shapes of $N_{P,i}(\eta_{P,i}^{(k)})$.
 It is possible to divide nodes into two classes, that differ in the behavior of the processing energy consumption against the compression ratio.
 
 \begin{itemize}
  \item $N_{P,i}(\eta_{P,i})$ is increasing and concave in $\eta_{P,i}$ (i.e., compressing more requires less energy). Two examples of this class are the Lightweight Temporal Compression (LTC)  algorithm and the Fourier-based Low Pass Filter (DCT-LPF) algorithm. Further, for these two techniques, function $N_{P,i}(\eta_{P,i})$ can be accurately approximated as linear: $N_{P,i}(\eta_{P,i}) = \alpha_{P,i}\,\eta_{P,i}  + \beta_{P,i}$ (we refer the reader to~\cite{zordan} for details). The corresponding energy consumption becomes (we recall that $\eta_{P,i}^{(k)} = L_{i}^{(k)}/L_{0,i}^{(k)}$):~
 \begin{align}
   E_{P,i}^{(k)}  = E_{0,i} \,\alpha_{P,i}\,\eta_{P,i}^{(k)}\,L_{0,i}^{(k)} + E_{0,i}\, L_{0,i}^{(k)} \,\beta_{P,i}.  \label{eq:e_processing_lin}
 \end{align}

 \noindent Notice that the second term is independent of the level of compression employed;
 
  \item $N_{P,i}(\eta_{P,i})$ decreases with $\eta_{P,i}$ (i.e., compressing more requires more energy). This case models compression algorithms which are more sophisticated than the previous ones, and will be analyzed in future work. 
 \end{itemize}

 In this work, we consider only the former class of algorithms. We note that our analytical results do not exploit the linearity of~\eqref{eq:e_processing_lin} but only assume $N_{P,i}(\eta_{P,i}^{(k)})$ to be increasing in $\eta_{P,i}^{(k)}$.

\emph{\textbf{Data transmission}.} If device $i$ transmits for a time interval of duration $\tau_i^{(k)}$ seconds with a constant power $P_i^{(k)}$, it consumes the following amount of energy:~
\begin{align}
 E_{\ttx,i}^{(k)} =  \frac{P_i^{(k)} \cdot \tau_i^{(k)}}{\eta_{A,i}},   \label{eq:e_tx}
\end{align}

  \noindent where the numerator represents the energy of the radio signal transmitted over the air, and $\eta_{A} \in (0,1]$ is a constant term that models the efficiency of the power amplifier of the antenna.

\emph{\textbf{Data sensing and circuitry costs}.}
We also account for the energy spent to sense the environment and the energy losses due to circuitry, such as the energy spent for node switches between sleep and active modes, the synchronization costs, and the additional energy required by transmission. 
We thus define the circuitry energy consumption as:~
\begin{align}
 E_{C,i}^{(k)} =  \beta_{{\rm S},i}^{(k)} + \beta_{C,i}^{(k)} + \mathcal{E}_{C,i} \cdot \tau_i^{(k)} = \beta_i^{(k)} + \mathcal{E}_{C,i} \cdot \tau_i^{(k)},  \label{eq:e_circuitry}
\end{align}

 \noindent where $\beta_{{\rm S},i}^{(k)}$ and $\beta_{C,i}^{(k)}$ represent the constant sensing and circuitry contributions, respectively, and $\mathcal{E}_{C,i}$ is the rate of circuitry energy consumption during data transmission.
Note that no energy is wasted because of collisions and overhearing, since we adopt a TDMA-based access mechanism and devices are granted exclusive use of the communication channel in their slot (a single frame is composed by $N$ slots).

\medskip
The total energy consumption of a node in frame $k$ is obtained by summing Equations~\eqref{eq:e_processing}-\eqref{eq:e_circuitry}:~
\begin{align} \label{eq:energy_used}
  E_{{\rm used},i}^{(k)}  & =E_{P,i}^{(k)} + E_{\ttx,i}^{(k)} + E_{C,i}^{(k)}. 
\end{align}


\subsection{Optimization Problem} \label{subsec:opt_problem}
Our goal is to determine a joint compression-transmission policy that simultaneously satisfies the QoS requirements and extends the network lifetime. To handle these two conflicting objectives, we set up the following weighted optimization problem:~
\begin{align}
    \min_{\textstyle \mathbf{E}^{(0)},\mathbf{E}^{(1)},\ldots} \sigma \frac{1}{n} \sum_{k = 1}^n f_{\rm FOP}^{(k)}(\mathbf{E}^{(k)}) - (1-\sigma)\,n, \label{eq:general_output}
\end{align}

\noindent where $\sigma \in [0,1]$ is a weighting factor,  and $n$ is the effective lifetime of the system, which we defined as the first frame in which at least one node dies, i.e., it does not have enough energy in its battery to transmit any more data with an acceptable distortion level. Notice that the network lifetime is an outcome of the energy assignment, and therefore we must consider $k \in \mathbb{N}$ for the optimization variables; this makes the optimization problem more challenging because the number of optimization variables is not known a priori. The first term of~\eqref{eq:general_output} is a function of the distortion, $f_{\rm FOP}^{(k)}(\mathbf{E}^{(k)})$, averaged over the lifetime $n$, while the second term is an arbitrary decreasing function of $n$, and is used to express the tradeoff between distortion and lifetime.
Function $f_{\rm FOP}^{(k)}(\mathbf{E}^{(k)})$ represents the distortion in slot $k$ given the energy vector $\mathbf{E}^{(k)}$, and is defined as follows:~
 \begin{align}
  f_{\rm FOP}^{(k)}(\mathbf{E}^{(k)}) = \min_{\textstyle \bm{\tau}^{(k)}, \mathbf{L}^{(k)}, \mathbf{P}^{(k)}} \, \max_{i \in \mathcal{N}} \frac{D_i^{(k)}}{D_{\tth, i}^{(k)}},
  \label{eq:FOP_minmax}
 \end{align}
 
\noindent and will be presented in its extended form in~\eqref{prob:FOP}, where also the dependence on $\mathbf{E}^{(k)}$ will be clarified. In brief, we aim at minimizing the normalized distortion of the ``worst'' user in slot $k$ in order to guarantee fairness. To achieve that, we must decide how much each node should compress its packet, and with what power and for how long it should transmit.

Note that, if the energy consumption $\mathbf{E}^{(k)}$ were given $\forall k$, the lifetime $n$ would be uniquely determined and the first part of~\eqref{eq:general_output} could be solved independently.
Based on this observation, we decided to structure the problem in a modular fashion. This choice introduces a level of independence between the building blocks, that can be slightly adapted to meet different requirements in a separate way while keeping the overall framework simple.
The two blocks have the following objectives. 
\begin{enumerate}
 \item \emph{Energy Allocation Problem} (EAP): it is the main problem, with the goal expressed in Equation~\eqref{eq:general_output}. It defines the energy allocation over time $\{\mathbf{E}^{(0)},\mathbf{E}^{(1)},\ldots\}$ and thus determines the effective lifetime $n$. EAP assumes to know the mapping between the function $f_{\rm FOP}^{(k)}$ and the energy vector $\mathbf{E}^{(k)}$;
 \item \emph{Frame-Oriented Problem} (FOP): it focuses on a single frame $k$, assuming that the energy vector $\mathbf{E}^{(k)}$ is given. FOP is completely unaware of the energy allocation in the other slots as well as of the network lifetime. Its goal is to determine the transmission durations and powers and the compression ratios that minimize Equation~\eqref{eq:FOP_minmax}.
\end{enumerate}

\begin{figure}[H]
  \centering
  \includegraphics[trim = 0mm 0mm 0mm 0mm,  clip, width=1\columnwidth]{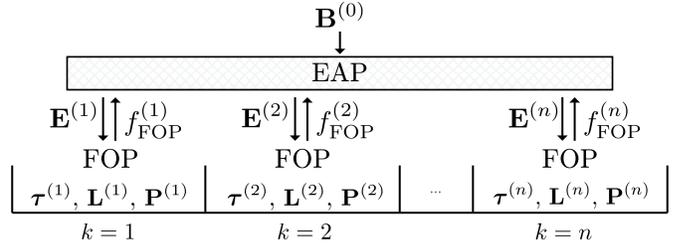}
  \caption{Structure of the problem: modules.}
  \label{fig:structure}
\end{figure}

\figurename~\ref{fig:structure} shows the relation between the blocks.
In practice, the two problems are tightly coupled: EAP defines the energy allocation to use in every slot, which is used by FOP to determine~\eqref{eq:FOP_minmax} and, on the other hand, the output of FOP influences EAP through~\eqref{eq:general_output}.
We developed two versions of FOP, which differ in the level of CSI available at the nodes.
In the next sections, we will discuss FOP and EAP, and how the two problems are interrelated.

\section{Frame-Oriented Problem (FOP)}\label{sec:FOP}

The Frame-Oriented Problem does not consider the time evolution of the network and operates at the frame level.  The  amount of energy available to each node is thus fixed. The goal of FOP is to determine the optimal compression-transmission policy that minimizes the maximum normalized distortion experienced by the users in a single frame $k$.

For each user $i\in\mathcal{N}$, FOP determines:~
\begin{enumerate}
 \item the degree of compression $\eta_{P,i}^{(k)}$, which is strictly related to the distortion and has an impact on the energy consumed for processing as well as for transmitting. This is equivalent to determining the size of the data to transmit, since $L_i^{(k)} =  \eta_{P,i}^{(k)}\,L_{0,i}^{(k)}$;
 \item the transmission power $P_i^{(k)}$, which influences the transmission rate and the transmission energy consumption;
 \item the transmission duration $\tau_i^{(k)}$, which relates $L_i^{(k)}$ to the channel rate $r_i^{(k)}$ and affects the transmission energy.
\end{enumerate}

Since FOP focuses on single frames, for ease of notation we will omit the dependence on the frame index $k$ throughout this section, if not misleading. In this case, boldface letters refer to column vectors that span over the $N$ users for the considered frame.

\subsection{Optimization Problem}\label{subsec:FOP_optimization} 

We set up FOP as a minimax problem, so as to guarantee fairness among users:~
\begin{subequations} \label{prob:FOP}
 \begin{flalign}
    \text{FOP:} && & \min_{\textstyle \bm{\tau}, \mathbf{L}, \mathbf{P}} \, \max_{i \in \mathcal{N}} \frac{D_i}{D_{\tth, i}}, & \label{eq:FOP_objective}
 \end{flalign}
 \vspace{-\belowdisplayskip}
 \vspace{-\abovedisplayskip}
 \begin{alignat}{2}
    \shortintertext{subject to:}
    & D_i =  b_i \left( {\left(\dfrac{L_{0,i}}{L_i}\right)}^{a_i} - 1 \right) \leq D_{\tth, i}, \quad && \forall i, \label{eq:FOP_distortion}\\
    & L_i \leq \tau_i \, r_i, \quad && \forall i, \label{eq:FOP_capacity_const} \\
    & E_{{\rm used},i} \leq E_i, \quad && \forall i, \label{eq:FOP_energy_const}\\
    & P_{{\rm min},i} \le P_i \le P_{{\rm max},i}, \quad && \forall i, \label{eq:FOP_power}\\
    &  \sum_{i = 1}^N \tau_i \le T. \label{eq:FOP_tau}
 \end{alignat}
\end{subequations}
\noindent The distortion was defined in~\eqref{eq:dist}, and Constraint~\eqref{eq:FOP_distortion} guarantees that it does not exceed the given threshold for any user. Notice that, since $D_i$ is a non-negative quantity, the packet size $L_i$ lies in $[0, L_{0,i}]$. Inequality~\eqref{eq:FOP_capacity_const} ensures not to exceed the channel capacity given in~\eqref{eq:rate}. 
 The energy constraint~\eqref{eq:FOP_energy_const} ensures that each node does not consume more energy than what is available (see~\eqref{eq:energy_used}). 
 Finally, Constraint~\eqref{eq:FOP_tau} limits the total time allocated to the users to the frame duration. Notice that this is the only constraint that considers the users jointly: without it, FOP could be readily decomposed into $N$ separate problems.

The optimization problem~\eqref{prob:FOP} and the concepts introduced hitherto have general validity and hold regardless of the knowledge of the channel status.
In the following, we first introduce the solution of~\eqref{prob:FOP} without fading (Section~\ref{subsec:FOP_full_CSI}), which is simpler to determine and can be used as a building block for the other case (Section~\ref{sec:FOP_avg_CSI}).

\subsection{Solution of FOP with Full CSI -- Policy Without Fading} \label{subsec:FOP_full_CSI}

In this scenario, we assume nodes to have full CSI, i.e., the gain $h_i$ of all channels is known exactly for all frames a priori. This is equivalent to a scenario without fading, since small-scale fading effects cannot be accurately tracked in advance.

Without loss of generality, we take Constraint~\eqref{eq:FOP_capacity_const} with equality, which is equivalent to choosing the smallest time possible when $L_i$ and $P_i$ are fixed. 
This implies that $\bf{L}$ can be expressed as a function of $\mathbf{P}$ and $\bm{\tau}$, and thus removed from the optimization variables.

The objective function~\eqref{eq:FOP_objective} can be equivalently formulated by introducing an auxiliary optimization variable $\Gamma$:~
\begin{subequations} \label{prob:FOP_gamma}
  \begin{flalign}
      \text{FOP$_\Gamma$:} && & \min_{\textstyle \Gamma, \bm{\tau}, \mathbf{P}} \Gamma,  & \label{eq:FOP_objective_gamma}
  \end{flalign}
  \vspace{-\belowdisplayskip}
  \vspace{-\abovedisplayskip}
  \begin{alignat}{2}
  \shortintertext{subject to:}
  & \frac{D_i}{D_{\tth, i}} \leq \Gamma, \quad  \forall i, \label{eq:FOP_gamma} \\
  & L_i = \tau_i \, r_i, \quad \forall i, \label{eq:FOP_capacity_const_equality} \\
  & \mbox{Constraints } \eqref{eq:FOP_distortion}-\eqref{eq:FOP_tau}.
  \end{alignat}
\end{subequations}

 
We denote the solution of \text{FOP$_\Gamma$} as $\Gamma^\star$ and, since it depends on the energy allocated to each user (see~\eqref{eq:FOP_energy_const}), we explicitely write $\Gamma^\star = f_{\rm FOP}(\mathbf{E})$. Note that only distortions below the threshold are acceptable (see~\eqref{eq:FOP_distortion}), which means that $\Gamma^\star\le 1$, otherwise FOP is infeasible.
In the remainder of this subsection, we propose an efficient technique to extract one optimal solution.

\begin{lemma}\label{lemma:FOP_gamma}
    In at least one optimal solution, Constraint~\eqref{eq:FOP_gamma} is satisfied with equality for every $i$.
    \begin{proof}
        See \appendixname~\ref{proof:FOP_gamma}.
    \end{proof}
\end{lemma}

Assume that $\Gamma$ is given; then, thanks to Lemma~\ref{lemma:FOP_gamma}, the distortion and the compression degree are fixed for each user. 
Being $\bf{L}$ fixed, the capacity constraint~\eqref{eq:FOP_capacity_const_equality} allows us to explicitly write $\tau_i$ as a function of $P_i$. Intuitively, we prefer high transmission powers, since $\tau_i$ and $P_i$ are inversely proportional to each other, and the allocation of shorter transmission times makes it more likely to satisfy the frame duration constraint~\eqref{eq:FOP_tau}. For this reason, we choose the highest $P_i$, namely $P_i^\star$, that satisfies both the energy~\eqref{eq:FOP_energy_const} and the power~\eqref{eq:FOP_power} constraints. By combining~\eqref{eq:FOP_energy_const} and~\eqref{eq:FOP_capacity_const_equality}, we obtain:~
\begin{subequations}
\label{prob:g}
  \begin{flalign}
  && & P_i^\star = \max_{P_i \in [P_{\min,i},P_{\max,i}]} P_i,  &
  \end{flalign}
  \vspace{-\belowdisplayskip}
  \vspace{-\abovedisplayskip}
  \begin{alignat}{2}
  \shortintertext{subject to:}
  &  g_i(P_i) \leq \frac{W}{L_i}(E_i - E_{P,i}(L_i)  + \beta_i), \label{eq:g_Pi}
  \end{alignat}
\end{subequations}

\noindent with $g_i(x) \triangleq (x/\eta_{A,i} + \mathcal{E}_{C,i})/\log_2(1+h_i\,x)$. We explicitely wrote $E_{P,i}(L_i)$ to highlight the dependence on $L_i$.

\begin{figure}[t]
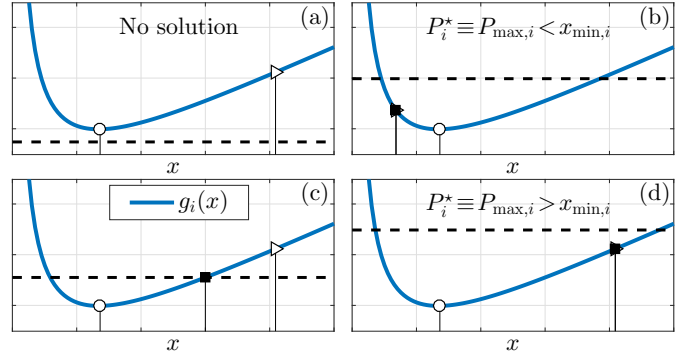

    \centering
    \begin{minipage}[t]{.49\columnwidth}
        \centering
        \includegraphics[trim = 0mm 0mm 0mm 0mm,  clip, width=1\columnwidth]{plot_g(a).eps}
    \end{minipage}%
    \hfill%
    \begin{minipage}[t]{.49\columnwidth}
        \centering
        \includegraphics[trim = 0mm 0mm 0mm 0mm,  clip, width=1\columnwidth]{plot_g(b).eps}
    \end{minipage}\\
    \begin{minipage}[t]{.49\columnwidth}
        \centering
        \includegraphics[trim = 0mm 0mm 0mm 0mm,  clip, width=1\columnwidth]{plot_g(c).eps}
    \end{minipage}%
    \hfill%
    \begin{minipage}[t]{.49\columnwidth}
        \centering
        \includegraphics[trim = 0mm 0mm 0mm 0mm,  clip, width=1\columnwidth]{plot_g(d).eps}
    \end{minipage}\\
    \hspace{.005\textwidth}
    \begin{minipage}[t]{1\columnwidth}
        \caption{Function $g_i(x)$ in four different cases when solving~\eqref{eq:g_Pi}. The dash-line represents different values of $W/L_i(E_i - E_{P,i}(L_i)  + \beta_i)$. The empty circle and triangle markers represent $x_{{\rm min},i}$ and $P_{{\rm max},i}$, respectively. The black square markers represent $P_i^\star = \overline{P}_i$.}
        \label{fig:plot_g}
    \end{minipage}
    \vspace{-.5cm}
\end{figure}

Note that all the terms on the right-hand side (RHS) of~\eqref{eq:g_Pi} are fixed: $E_i$ is given, $L_i$ is derived from $\Gamma$ through Lemma~\ref{lemma:FOP_gamma}, and the remaining are system parameters. It can be shown that $g_i(x)$ is a decreasing-increasing function of~$x$ and that it admits only one minimum, as shown in \figurename~\ref{fig:plot_g}. 
To solve~\eqref{prob:g}, we first use the golden search algorithm to find the point of minimum $x_{\min,i}$, which is then used to determine the amount of power $\overline{P}_i$ that solves~\eqref{prob:g} when the constraint on $P_{\min,i}$ is neglected. If $g(x_{\min,i}) > W/L_i(E_i - E_{P,i}(L_i)  + \beta_i)$, no solution exists and $\overline{P}_i$ is undefined (case \emph{a} of \figurename~\ref{fig:plot_g}). Otherwise, either $x_{\min,i} \geq P_{\max,i}$,  and then we set $\overline{P}_i = P_{\max,i}$ (case \emph{b}); or $x_{\min,i} < P_{\max,i}$ (case \emph{c}), and in this latter case we use a dichotomic search in $[x_{\min,i},P_{\max,i}]$ to find $\overline{P}_i$ as the intersection between $W/L_i(E_i - E_{P,i}(L_i)  + \beta_i)$ and $g(P_i)$, i.e., the value of $P_i$ that satisfies~\eqref{eq:g_Pi} with equality. It this is not possible, we take $\overline{P}_i = P_{\max,i}$ (case \emph{d}). 

\noindent If, using the previous procedure, $\overline{P}_i$ does not exist or $\overline{P}_i < P_{{\rm min},i}$, then $P_i^\star$ is not defined and the problem is infeasible for the given $\Gamma$; otherwise, it coincides with $\overline{P}_i$.

In practice, for a fixed $\Gamma$, we are able to determine the compression level for each user, their transmission powers and the duration of their transmission intervals. The following result allows us to determine the optimal $\Gamma^\star$.

\begin{thm}\label{thm:Gamma1_Gamma2}
    If \emph{FOP}$_\Gamma$ is feasible for a fixed $\Gamma^\prime \le 1$, then \emph{FOP}$_\Gamma$ is feasible for all $\Gamma^{\prime\prime}$ such that $\Gamma^\prime \leq \Gamma^{\prime\prime} \leq 1$.
    \begin{proof}
        See \appendixname~\ref{proof:Gamma1_Gamma2}.
    \end{proof}
\end{thm}

According to~\eqref{eq:FOP_objective_gamma}, $\Gamma^\star$ is the minimum value of $\Gamma$ which is feasible and can be found using the following corollary.

\begin{corol} \label{corol:Gamma_star}
    $\Gamma^\star$ can be found with a bisection search over the interval $[0, 1]$.
\end{corol}

Few operations are sufficient to find $\Gamma^\star$ with a very high precision, from which the optimal compression ratios are derived straightforwardly, and the other optimization variables, $\bm{\tau}$ and $\textbf{P}$, are determined as described previously.

\section{FOP with Statistical CSI -- Policy With Fading} \label{sec:FOP_avg_CSI}

In this section, we explicitly consider the fading contributions and design a policy to determine the transmission power and duration assuming to have only a statistical knowledge of the channel. Again, we will omit the dependence on the frame index throughout this subsection, if not ambiguous.
In particular, the channel coefficient can be decomposed as $h_i = h_{0,i}\cdot \theta_i$, where $h_{0,i}$ represents the average channel gain given by path loss and shadowing, and $\theta_i$ is the realization of a random variable $\Theta_i$ that models the fast-fading effects. Function $f_{\Theta_i}(\theta_i)$ represents the probability density function of $\Theta_i$ (e.g., $\Theta_i \sim {\rm Exp}(1)$ for Rayleigh fading). 
To compute the policy, we assume that $h_{0,i}$ is known for future frames, whereas $\theta_i$ is known only in the current slot.

First, we reformulate the definition of distortion of user $i$ in order to include the statistical knowledge of the fading realization (throughout this subsection, and when we deal with the statistical CSI case in general, we substitute $D_i$ in~\eqref{prob:FOP} with $\overline{D}_i$):\footnote{Definition~\eqref{eq:D_avg} degenerates in $D_i$ when fading is not considered (i.e., $f_{\Theta_i}(\theta_i)$ is a Dirac delta function) and if $\theta_{\ttx,i}$ is sufficiently low.}~
\begin{align} \label{eq:D_avg}
    \overline{D}_i \triangleq \ddfrac{\int_{\theta_{\ttx,i}}^\infty \delta_i(\theta_i,\tau_i,\rho_i(\theta_i)) f_{\Theta_i}(\theta_i)\ \mbox{d}\theta_i}{\int_{\theta_{\ttx,i}}^\infty f_{\Theta_i}(\theta_i)\ \mbox{d}\theta_i}.
\end{align}

\noindent The fading value $\theta_{\ttx,i}$ guarantees that a transmission is performed only if the corresponding channel gain is sufficiently high, and will be further discussed in the remainder of this section. Accordingly, the denominator is a normalization term that represents the probability that node $i$ transmits in the considered frame. The distortion function
$\delta_i$ is defined by combining~\eqref{eq:dist},~\eqref{eq:rate}, and Constraint~\eqref{eq:FOP_capacity_const} taken with equality:~ 
\begin{align} \label{eq:delta_funct}
    \delta_i(\theta_i,\tau_i,P_i) \!=\! \left[ b_i\left(\dfrac{L_{0,i}}{W \tau_i \log_2(1+h_{0,i}\theta_i P_i)}\right)^{a_i} \!\!\!\!- 1\right]^+. 
\end{align}

\noindent Finally, $\rho_i :  \mathbb{R}^+ \to [P_{\min,i},P_{\max,i}]$ is the transmission power function, that associates a fading realization with the corresponding transmission power. Indeed, differently from Section~\ref{subsec:FOP_full_CSI}, where we allocated a unique transmission power value $P_i$ for each user, choosing a unique $P_i$ for a whole frame may be suboptimal when the channel is affected by fading. In this case, a dynamic approach that adapts $P_i$ to $\theta_i$ is required.
Thus, for every frame, the optimization problem decides the function $\rho_i(\theta_i)$ rather than a single value $P_i$.\footnote{Note that we could have also adapted the transmission duration $\tau_i$ according to $\theta_i$, in addition to $P_i$. However, the transmission durations of the $N$ users are tightly coupled (see~\eqref{eq:FOP_tau}) and the use of adaptive $\tau_i$ may lead to collisions if multiple devices transmit simultaneously, and also to troublesome coordination issues, since, in the current frame, each node executes the policy on-board and independently of the others.} Then, when actually transmitting, node $i$ examines its fading coefficient $\theta_i$, and picks the corresponding transmission power using $\rho_i(\theta_i)$. We remark that, also in this case, the power is kept constant during a transmission.

In the following, we first describe how to define $\theta_{\ttx,i}$, then we introduce function $\rho_i(\cdot)$, and finally we use it to optimally find $\tau_i$.

\subsection{Transmission Probability} 
When fading is taken into account, the channel condition may be too bad and lead to an unacceptable level of distortion. To avoid wastage of resources in useless transmissions, we do not allow a node to send its data if the channel coefficient is larger than a certain threshold $\theta_{\ttx,i}$ (a device looks at its channel realization $\theta_i$ and compares it with $\theta_{\ttx,i}$ to decide whether to transmit or not). Consequently, the probability of accessing the channel (i.e., $\int_{\theta_{\ttx,i}}^\infty f_{\Theta_i}(\theta_i) \ \mbox{d}\theta_i$) in a frame is in general smaller than one.
Clearly, if $\theta_{\ttx,i}$ were very high, the transmitted packets would always have a very low distortion, but the probability of transmitting would be small. Thus, there exists a tradeoff between transmission probability and quality of the transmitted data. In order to guarantee a certain level of QoS, we fix the transmission probability $\overline{\rm Pr}_{\ttx,i}$ a priori, and derive $\theta_{\ttx,i}$ by imposing:~
\begin{align} \label{eq:Pr_tx}
    \int_{\theta_{\ttx,i}}^\infty f_{\Theta_i}(\theta_i) \ \mbox{d}\theta_i = \overline{\rm Pr}_{\ttx,i}.
\end{align}

\noindent For example, with Rayleigh fading, it is $\theta_{\ttx,i} = -\log(\overline{\rm Pr}_{\ttx,i})$.

\subsection{Optimal Power Allocation} 

When the channel fading is $\theta_i$ and the transmission duration $\tau_i$ is given, the transmission power $\rho_i(\theta_i)$ must satisfy the energy constraint~\eqref{eq:FOP_energy_const} and be as large as possible (in order to maximize the amount of transmitted data:\footnote{Problems~\eqref{prob:g} and~\eqref{prob:g_theta} are equivalent. The only difference is that in~\eqref{prob:g} the packet size $L_i$ is fixed, whereas in~\eqref{prob:g_theta} it is the transmission duration $\tau_i$ to be fixed.}~
\begin{subequations}
\label{prob:g_theta}
  \begin{flalign}
     && & \rho_i(\theta_i) = \max_{P_i \in [P_{\min,i},P_{\max,i}]} P_i,  &
  \end{flalign}
  \vspace{-\belowdisplayskip}
  \vspace{-\abovedisplayskip}
  \begin{alignat}{2}
  \shortintertext{subject to:}
  &  \begin{split}  \label{eq:fading_EC_less_Ei}
        E_{{\rm P},i}(L_i)  + \beta_i + \bigg(\frac{P_i}{\eta_{A,i}} + \mathcal{E}_{C,i}\bigg) \tau_i &\leq E_i
    \end{split}
  \end{alignat}
\end{subequations}

\noindent Notice that $E_{{\rm P},i}(L_i)$ does depend on the transmission power through the packet size, as in~\eqref{eq:FOP_capacity_const}, and, in particular, it is increasing in $P_i$. Therefore, since the left-hand side (LHS) of~\eqref{eq:fading_EC_less_Ei} increases with $P_i$, we can use a bisection search to determine the quantity $\smash{\overline{P}}_i^{\theta}$, defined as the value of $P_i$ that satisfies~\eqref{eq:fading_EC_less_Ei} with equality. If $\smash{\overline{P}}_i^{\theta} \geq P_{\min,i}$, the optimal solution  is $\rho_i(\theta) = \min\{P_{\max,i},\,\smash{\overline{P}}_i^{\theta}\}$. Otherwise, no solution exists and this can be due to two different reasons.~
\begin{itemize}
    \item $\beta_i + ({P_{\min,i}}/{\eta_{A,i}} + \mathcal{E}_{C,i})\, \tau_i > E_i$, which means that the energy constraint cannot be satisfied even when using $L_i = 0$ in~\eqref{eq:fading_EC_less_Ei}. This happens because the transmission duration $\tau_i$ is too large and has to be reduced (see Section~\ref{subsec:opt_tau});
    \item Otherwise, the energy constraint can be satisfied by using $P_i = P_{\min,i}$, obtaining a compressed packet size $L_i > 0$. However, by imposing to use the whole channel capacity (i..e, $L_i = \tau_i \, r_i$), Constraint~\eqref{eq:fading_EC_less_Ei} would be violated; this means that $\theta_i$ and consequently $r_i$ are large (i.e., good channel conditions). In this case, we can underuse the channel and send fewer bits, allowing to perform the transmission. In particular, we introduce quantity $L_{\een,i}$ as:~
    \begin{align} \label{eq:L_en}
        L_{\een,i} \triangleq E_{P,i}^{-1}(E_i - \beta_i - (P_{\min,i}/{\eta_{A,i}} + \mathcal{E}_{C,i})\tau_i),
    \end{align}
    
    \noindent where $E_{P,i}^{-1}(\cdot)$ is the inverse function of $E_{P,i}(\cdot)$. In practice, the transmission is done with $\rho_i(\theta_i) = P_{\min,i}$ but, instead of transmitting $W \tau_i \log_2(1+h_{0,i}\theta_i P_{\min,i})$ bits (which is not possible since $\smash{\overline{P}}_i^{\theta} < P_{\min,i}$), we reduce the transmission rate (the capacity constraint~\eqref{eq:FOP_capacity_const} is satisfied with strict inequality) and send only $L_{\een,i}$ bits. By doing so, we satisfy both the energy constraint (with equality) and the power constraints. Notice that, in this case, the definition of distortion function in~\eqref{eq:delta_funct} must be adapted to account for $L_{\een,i}$ bits because Constraint~\eqref{eq:FOP_capacity_const} is not taken with equality.
\end{itemize}

\subsection{Optimal Transmission Duration}\label{subsec:opt_tau}

Given $\tau_i$, we are able to evaluate $\overline{D}_i$ by defining $\rho_i(\theta_i)$ and $\theta_{\ttx,i}$ as described previously. 
We also highlight that the distortion function $\delta_i(\cdot)$ decreases with $\theta_i$, as expected:~
\begin{align}
    \delta_i\big(\theta_i^\prime,\tau_i,\rho_i(\theta_i^\prime)\big) \geq \delta_i\big(\theta_i^{\prime\prime},\tau_i,\rho_i(\theta_i^{\prime\prime})\big), \quad \theta_i^\prime \leq \theta_i^{\prime\prime},
\end{align}

\noindent and in particular $\delta_i\big(\theta_{\ttx,i},\tau_i,\rho_i(\theta_{\ttx,i})\big) \geq \delta_i\big(\theta_i,\tau_i,\rho_i(\theta_i)\big)$ for every $\theta_i \geq \theta_{\ttx,i}$. Therefore, if~
\begin{align} \label{eq:delta_th}
    \delta_i\big(\theta_{\ttx,i},\tau_i,\rho_i(\theta_{\ttx,i})\big) \leq D_{\tth,i}
\end{align}

\noindent is satisfied, a similar condition holds for every $\theta_i \geq \theta_{\ttx,i}$, and we are guaranteed to transmit packets with acceptable distortion levels. If~\eqref{eq:delta_th} is instead violated, the chosen $\tau_i$ is not an acceptable solution, because it induces transmissions with too large distortions. We now discuss how to determine the transmission duration in order to meet~\eqref{eq:delta_th}. First, we prove an important property of the distortion function.

\begin{lemma} \label{lemma:delta_tau}
    The distortion function $\delta_i(\theta_i,\tau_i,\rho_i(\theta_i))$, decreases with $\tau_i$ until a minimum distortion point at $\tau_i^{\min}$, and increases for $\tau_i > \tau_i^{\min}$.
    \begin{proof}
        See \appendixname~\ref{proof:delta_tau}.
    \end{proof}
\end{lemma}

Therefore, if $\tau_i^{\min}$ computed at $\theta_{\ttx,i}$ satisfies~\eqref{eq:delta_th}, there exists an interval $[\tau_i^{\rm low},\tau_i^{\rm high}]$, with $\tau_i^{\rm low} \leq \tau_i^{\min} \leq \tau_i^{\rm high}$, for which~\eqref{eq:delta_th} still holds. Otherwise, it is not possible to solve FOP. 
We remark that the average distortion $\overline{D}_i$ is the sum (integral) of many $\delta_i(\theta_i,\tau_i,\rho_i(\theta_i))$ functions that exhibit a minimum at $\tau_i^{\min}$, as shown in Lemma~\ref{lemma:delta_tau}. However, $\tau_i^{\min}$ does depend on $\theta_i$; thus, in general $\overline{D}_i$ does not inherit the properties of $\delta_i(\cdot)$ (Lemma~\ref{lemma:delta_tau}) and may have multiple points of minimum as a function of $\tau_i$. 

\medskip

Provided that we can satisfy~\eqref{eq:delta_th}, and since our goal is to solve FOP (Problem~\eqref{prob:FOP}), we then proceed as in Section~\ref{subsec:FOP_full_CSI} and convert FOP to its equivalent form FOP$_\Gamma$.
To derive the optimal solution, namely $\tau_i^\star$, we hence look at the \emph{smallest} value of $\tau_i \geq \tau_i^{\rm low}$ that satisfies $\overline{D}_i \leq \Gamma D_{\tth,i}$, for a fixed $\Gamma$. Indeed, using the smallest transmission duration guarantees to satisfy Constraint~\eqref{eq:FOP_tau}, if this is possible.\footnote{Because of the way in which the optimal $\tau_i$ is defined, Theorem~\ref{thm:Gamma1_Gamma2}, and consequently Corollary~\ref{corol:Gamma_star} still hold in the statistical CSI case.} 
Finally, the optimal $\Gamma^\star$ is obtained as in Corollary~\ref{corol:Gamma_star}.

\medskip

In summary, the procedure proposed in this section explicitly accounts for unknown channel status by introducing the notion of ``expected distortion.'' A device transmits only if the channel gain is sufficiently high ($\theta_i \geq \theta_{\ttx,i}$), since a deep fade would lead to unacceptable data quality. To find $\theta_{\ttx,i}$, we defined the probability of performing a transmission, which is fixed a priori. Then, the optimal $\Gamma^\star$ is determined using a bisection search; for every fixed $\Gamma$, we used a dynamic power allocation policy which adapts the $P_i$ to the channel status, and we found $\tau_i$ as the smallest value that makes the problem feasible for every $\theta_i \geq \theta_{\ttx,i}$.

\section{Notes on FOP} \label{sec:FOP_notes}
In the previous sections, we described how to solve FOP with different knowledge of the status of the channels, but we did not consider them jointly. Also, we identified the conditions under which FOP is not feasible, but did not mention how to manage such situations. Further, we recall that EAP relies on the output of FOP (see~\eqref{eq:general_output}), hence it would be desirable to have a mathematical characterization of $f_{\rm FOP}^{(k)}$.
In this section, we handle all these problems.


\subsection{Sub-Optimal Policy with Fading} \label{subsec:sub_opt_pol}

The policy of Section~\ref{subsec:FOP_full_CSI} cannot be applied to a fading scenario directly, since we cannot predict the channel evolution. Therefore, the only possible choice to use it when fading is taken into account is to impose $\theta_i = \theta_{\ttx,i}$ and derive the policy accordingly. Indeed, if the solution is feasible at $\theta_{\ttx,i}$, it is guaranteed to be feasible even for better channel conditions. Therefore, the policy with full CSI may still be used in more realistic scenarios with fading, but it will have worse performance than the optimal version of Section~\ref{sec:FOP_avg_CSI}, as we will discuss in the numerical evaluation.

\subsection{Infeasibility of FOP: Dismission Policy} \label{subsec:dismission}

 In both scenarios, it may happen that the optimization over all $N$ users fails and FOP turns out to be infeasible in a specific frame. This happens if at least one constraint of FOP is not satisfied, i.e., there exists no allocation of $\bm{\tau}$, $\mathbf{L}$, and $\mathbf{P}$ that allows all users to transmit their packets in the considered frame with the allocated energy. Notice that FOP is infeasible even if the time, capacity, power, and energy constraints can be met but at least one user exceeds its threshold distortion, thereby violating the QoS constraint~\eqref{eq:FOP_distortion}.

Since we do not allow any constraint to be relaxed, the only strategies available when FOP is infeasible are: allocating a larger amount of energy, dismissing some users, or choosing a longer frame duration. 
In this paper, we study the first two strategies and propose a technique that leverages on both of them. 
We now discuss how and  when to dismiss users, whereas the energy allocation problem will be analyzed in Section~\ref{sec:EAP}.


Let $\underline{\tau}_i$ be the minimum transmission time user $i$ needs to transmit its data in the considered frame. This value is obtained when $P_i = P_{\max,i}$ and the distortion is the maximum allowed, i.e., $ D_i = D_{\tth,i}$, which means that the size of the packet to send, $L_i$, is the smallest possible. Then, we analyze the following inequality:~
\begin{align}
    \sum_{i=1}^N \underline{\tau}_i \leq T. \label{eq:t_min}
\end{align}

\noindent If the previous condition is not verified, it is impossible to allow all users to transmit in the same frame and satisfy their requirements at the same time, whatever the energy they have available. 
In this case, we assign different importance levels to the users and develop a priority-based dismission policy.\footnote{In particular, we consider \emph{fixed} priorities associated to users, and adopt a stochastic approach in which a higher admission priority entails a smaller probability of being dismissed.} When a user with low priority is dismissed, Condition~\eqref{eq:t_min} is rechecked, and the same dismission procedure is applied, if necessary.

We remark that this dismission policy is \emph{independent} of the energy allocation and is perfomed before EAP. 
 

\subsection{Structure of $f_{\rm FOP}^{(k)}(\cdot)$} \label{subsec:FOP_convex}

The solution for the energy allocation problem derived in the next section assumes that $f_{\rm FOP}^{(k)}(\mathbf{E}^{(k)})$ is \emph{decreasing} and \emph{convex} in $\mathbf{E}^{(k)}$. Although it is trivial to show that $f_{\rm FOP}^{(k)}(\mathbf{E}^{(k)})$ decreases with $\mathbf{E}^{(k)}$ (i.e., using more energy leads to lower distortions as all the terms in~\eqref{eq:energy_used} decrease with $L_i^{(k)}$), formally proving the convexity property is not easy in general (a simpler proof for the low SNR regime can be found in~\cite{Pielli2016}). Anyway, since from all our numerical results this property seems to hold, for the numerical evaluations we approximated $f_{\rm FOP}^{(k)}(\mathbf{E}^{(k)})$ with a convex function in order to guarantee that EAP is always solved correctly.

\section{Energy-Allocation Problem (EAP)} \label{sec:EAP}

So far, we have discussed how to maximize the quality of the transmitted information assuming there is a known amount of energy usable by each user. 
To solve the problem in~\eqref{eq:general_output}, it is necessary to distribute over time the energy available to each user, ${\bf B}^{(0)}$, balancing between lifetime and distortion according to the weight $\sigma$. In this section we introduce EAP and discuss how to solve~\eqref{eq:general_output}, where $f_{\rm FOP}^{(k)}$ is derived as described in the previous sections.

\subsection{Optimization Problem} \label{sec:EAP_optimization}
The solution of FOP cannot be expressed in close form, making it very challenging to directly relate FOP and the energy allocation.
We define EAP as the problem of allocating energy when the lifetime $n$ is fixed, and then exploit it to solve the general problem.

Given $n$, the optimization problem~\eqref{eq:general_output} reduces to:~
\begin{subequations} \label{prob:EAP}
\begin{flalign} \label{eq:EAP_objective}
    \text{EAP:} && & D_{\rm mean} \triangleq \min_{\textstyle \mathbf{E} } \;\dfrac{1}{n} \sum_{k = 1}^n f_{\rm FOP}^{(k)}(\mathbf{E}^{(k)}), &
\end{flalign}
\vspace{-\belowdisplayskip}
\vspace{-\abovedisplayskip}
\begin{alignat}{2}
\shortintertext{subject to:}
 & E_i^{(k)} \leq B_i^{(0)} - \sum_{j = 1}^{k-1} E_i^{(j)}, \quad && \forall i,\quad \forall k, \label{eq:EAP_batteries}\\
 & f_{\rm FOP}^{(k)}(\mathbf{E}^{(k)}) \mbox{ is feasible}, \quad && \forall k. \label{eq:EAP_feasible}
\end{alignat}
\end{subequations}

\noindent The objective function~\eqref{eq:EAP_objective} represents the average over the lifetime of the distortion metric in every frame, namely $f_{\rm FOP}^{(k)}(\cdot)$, which is convex in its argument $\mathbf{E}^{(k)}$ according to Section~\ref{subsec:FOP_convex}. The distortion function is computed by FOP, that thus needs to be feasible for all frames, as required by~\eqref{eq:EAP_feasible}, otherwise EAP is infeasible for the chosen $n$ (see Section~\ref{subsec:dismission}).  Constraint~\eqref{eq:EAP_batteries} specifies that the total energy assigned to node $i$ during the network lifetime cannot exceed the initial content of its battery, $B_i^{(0)}$. It can be rewritten in equivalent but simpler form as:~
\begin{align}
    \sum_{k = 1}^{n} E_i^{(k)} \leq B_i^{(0)}, \quad \forall i.
\end{align}

Notice that the size of the optimization variable $\mathbf{E}$ is known, as $n$ is fixed, and that the constraints induce a convex feasibility set because $f_{\rm FOP}^{(k)}(\mathbf{E}^{(k)})$ is convex in all the entries of $\mathbf{E}^{(k)}$.
This entails that EAP is a convex optimization problem and we leverage on this property to solve it.

\subsection{Solution of EAP}

We introduce a ``reduced'' version of EAP, where the optimization is done only over the energy vector $\mathbf{E}_\ell = [E_\ell^{(1)},\ldots,E_\ell^{(n)}],\, \ell \in \mathcal{N}$, and all the other variables in $\mathbf{E}$ are kept fixed:~
\begin{subequations} \label{prob:EAP_red}
\begin{flalign} \label{eq:EAP_red_objective}
    \text{EAP$_\ell$:} && & \min_{\textstyle \mathbf{E}_\ell} \sum_{k = 1}^n f_{\rm FOP}^{(k)}( \mathbf{E}^{(k)}), &
\end{flalign}
\vspace{-\belowdisplayskip}
\vspace{-\abovedisplayskip}
\begin{alignat}{2}
\shortintertext{subject to:}
 & \sum_{k = 1}^{n} E_\ell^{(k)} \leq B_\ell^{(0)}, \label{eq:EAP_red_batteries}\\
 & \underline{E}_\ell^{(k)} \leq E_\ell^{(k)} \leq \overline{E}_\ell^{(k)}, \quad && \forall k. \label{eq:EAP_red_feasible}
\end{alignat}
\end{subequations}

\noindent The objective~\eqref{eq:EAP_red_objective} and the energy constraint~\eqref{eq:EAP_red_batteries} are derived straightforwardly from~\eqref{eq:EAP_objective} and~\eqref{eq:EAP_batteries}, respectively. Constraint~\eqref{eq:EAP_red_feasible} ensures that the solution is feasible. In fact,
$\underline{E}_\ell^{(k)}$ represents the minimum amount of energy that node $\ell$ can use so that FOP is feasible in frame $k$ when the energy allocated to the other devices is $E_i^{(k)}, \, i\in \mathcal{N} \setminus \{\ell\}$. Indeed, if $E_\ell^{(k)}$ were too low, the distortion constraint~\eqref{eq:FOP_distortion} or the time constraint~\eqref{eq:FOP_tau} of FOP could not be satisfied. Similarly, $\smash{\overline{E}}_\ell^{(k)}$upper limits the energy available to node $\ell$ in frame $k$, because any $E_\ell^{(k)} \geq \smash{\overline{E}}_\ell^{(k)}$ does not make the objective function decrease further. 
Clearly, both $\underline{E}_\ell^{(k)}$ and $\smash{\overline{E}}_\ell^{(k)}$ strictly depend on the energy levels of the other nodes, $\{E_1^{(k)},\ldots,E_{\ell-1}^{(k)},E_{\ell+1}^{(k)},\ldots,E_N^{(k)}\}$. As discussed in Section~\ref{subsec:FOP_convex}, the solution of FOP for frame $k$ is decreasing in $E_\ell^{(k)}$ and, in particular, it turns out to be strictly decreasing (and convex) in $(\underline{E}_\ell^{(k)},\smash{\overline{E}}_\ell^{(k)})$. Consequently, EAP$_\ell$ can be solved in the dual domain by using the Lagrangian:~
\begin{subequations} \label{prob:EAP_red_L}
\begin{flalign} \label{eq:EAP_red_L_objective}
    && & \max_{\textstyle \lambda, \mathbf{E}_\ell} \;\sum_{k = 1}^n f_{\rm FOP}^{(k)}(\mathbf{E}^{(k)}) - \lambda \Big(\sum_{k = 1}^{n} E_\ell^{(k)} - B_\ell^{(0)} \Big)\!, &
\end{flalign}
\vspace{-\belowdisplayskip}
\vspace{-\abovedisplayskip}
\begin{alignat}{2}
\shortintertext{subject to:}
 & \lambda \geq 0, \\
 & \underline{E}_\ell^{(k)} \leq E_\ell^{(k)} \leq \overline{E}_\ell^{(k)}, \quad && \forall k. \label{eq:EAP_red_L_feasible}
\end{alignat}
\end{subequations}

The optimal solution must satisfy the Karush–Kuhn–Tucker conditions:~
\begin{align}
    &E_\ell^{(k)} = \max\{\underline{E}_\ell^{(k)},\min\{\overline{E}_\ell^{(k)},\psi^{-1}(\lambda)\} \}, \label{eq:E_ell_L}\\
    &\psi(E_\ell^{(k)}) \triangleq \frac{\partial f_{\rm FOP}^{(k)}(\mathbf{E}^{(k)})}{\partial E_\ell^{(k)}}, \label{eq:theta_derivative}
\end{align}

\noindent where $\lambda$ is such that $\sum_{k = 1}^{n} E_\ell^{(k)} = B_\ell^{(0)}$. Equation~\eqref{eq:E_ell_L} can be interpreted as a water-filling solution with minimum and maximum levels and in which the water level (i.e., the allocated energy) $\psi^{-1}(\lambda)$ is put in every slot $k$, if possible.

\emph{\textbf{Random Alternate Optimization.}} Based on EAP$_\ell$ that focuses on the optimization of one user at a time, we propose an alternate approach to solve EAP. In particular, we use Algorithm~\ref{alg:AO} to solve the general problem.
The key idea is to perform the optimization of a single user at every iteration, until the average of the distortion function over the lifetime, $D_{\rm mean} = 1/n\, \sum_{k = 1}^n f_{\rm FOP}^{(k)}(\mathbf{E}^{(k)})$, converges. The alternate optimization is done in Lines~\ref{alg:line:begin_AO}-\ref{alg:line:end_AO}: matrix $\mathbf{E}$ is used in Line~\ref{alg:line:EAP_ell} to solve EAP$_\ell$ (through~\eqref{prob:EAP_red_L}) for a specific user $\ell$ and update the $\ell$-th row of the energy matrix. It may happen that part of the initial energy is not used, i.e., $\sum_{k = 1}^{n} E_\ell^{(k)} < B_\ell^{(0)}$. Accordingly, Lines~\ref{alg:line:v_vect}-\ref{alg:line:end_AO} randomly distribute the residual energy $B_\ell^{(0)}-\sum_{k = 1}^{n} E_\ell^{(k)}$ among all the frames where $E_\ell^{(k)}$ is equal to $\overline{E}^{(k)}$ ($\chi\{\cdot\}$ is the indicator function). Note that, because of how $\overline{E}^{(k)}$ is defined, this operation does not change the distortion level obtained by solving EAP$_\ell$, but simply provides a new $\mathbf{E}_\ell$ that allows the alternate optimization to converge. 

We have the following key result. 
 
\begin{algorithm}[!t]
\caption{Random Alternate Optimization}\label{alg:AO}
\begin{algorithmic}[1]
\State Initialize a feasible $\mathbf{E}$
\State $D_{\rm mean} \gets \infty$
\While {$D_{\rm mean}$ has not converged}
\For {$\ell = 1,\ldots,N$} \label{alg:line:begin_AO}
\State $\mathbf{E}_\ell \gets $ solve EAP$_\ell(\mathbf{E})$ \label{alg:line:EAP_ell}
\State $v \gets$ prob. vector of size $\sum_k \chi\{E_\ell^{(k)} = \overline{E}^{(k)}\}$ \label{alg:line:v_vect}
\State $S \gets \sum_{k = 1}^{n} E_\ell^{(k)}$ \Comment consumed energy \label{alg:line:S}
\State $v_{\rm ind} \gets 1$ \Comment index of frames with $E_\ell^{(k)} = \overline{E}^{(k)}$
\For {$k = 1,\ldots,n$ \textbf{ such that } $E_\ell^{(k)} = \overline{E}^{(k)}$}
\State $E_\ell^{(k)} \gets v(v_{\rm ind})\cdot (B_\ell^{(0)}-S)$
\State $v_{\rm ind} \gets v_{\rm ind} + 1$  \label{alg:line:end_AO}
\EndFor
\EndFor
\State $D_{\rm mean} \gets 1/n \;\sum_{k = 1}^n f_{\rm FOP}^{(k)}(\mathbf{E}^{(k)})$
\EndWhile
\end{algorithmic}
\end{algorithm}

\begin{thm}\label{thm:alternate_opt}
    The alternate optimization approach of Algorithm~\ref{alg:AO}, in which only user $\ell$ is considered in a single step, leads to the optimal solution.
    \begin{proof}
        See \appendixname~\ref{proof:alternate_opt}.
    \end{proof}
\end{thm}

In summary, EAP optimally allocates energy over time using an alternate procedure: at every iteration of Algorithm~\ref{alg:AO}, EAP$_\ell$ is solved for the current node $\ell$, and FOP is invoked multiple times to evaluate the derivative in~\eqref{eq:theta_derivative}, which relates the allocated energy to the corresponding distortion metric.
Since this procedure holds for a specific lifetime, it is iterated over all possible values of $n$ and, by tuning $\sigma$, the network designer can choose a point in the tradeoff between lifetime and QoS.


\section{Numerical Evaluation}\label{sec:numerical_evaluation}

\begin{figure*}[t]
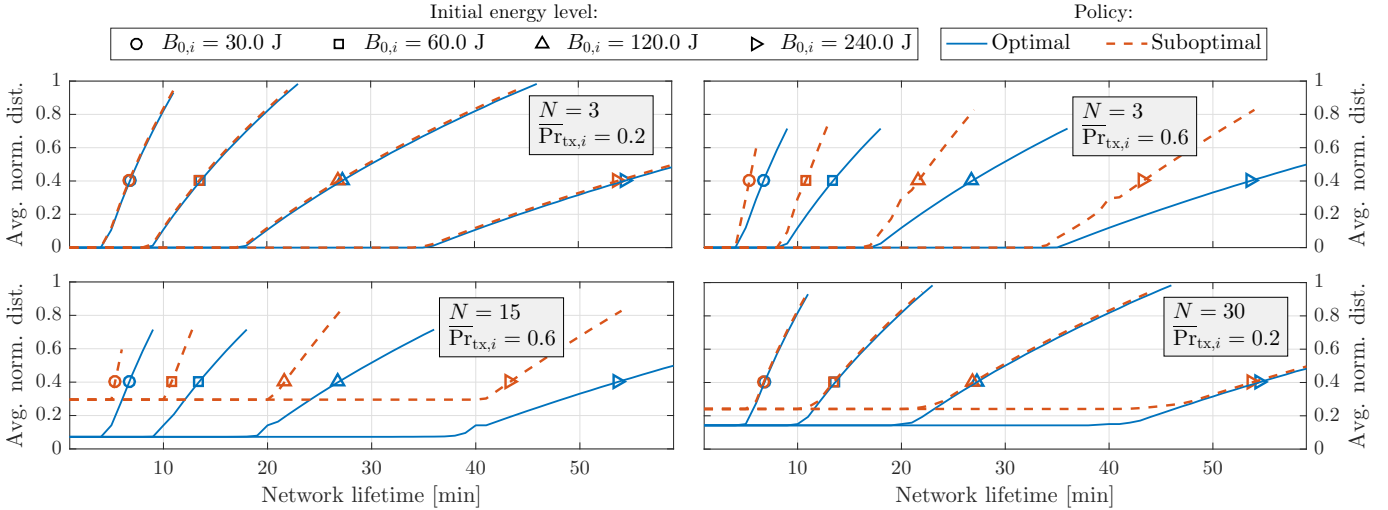

    \centering
    \begin{minipage}[b]{.85\textwidth}
        \centering
        \includegraphics[trim = 0mm 0mm 0mm 0mm,  clip, width=1\columnwidth]{symmetric_legend.eps}
    \end{minipage}\\
    \vspace{.1cm}
    \centering
    \begin{minipage}[b]{.49\textwidth}
        \centering
        \includegraphics[trim = 0mm 0mm 0mm 0mm,  clip, width=1\columnwidth]{symmetric_N3_Pr0v2.eps}
    \end{minipage}%
    \hfill%
    \begin{minipage}[b]{.49\textwidth}
        \centering
        \includegraphics[trim = 0mm 0mm 0mm 0mm,  clip, width=1\columnwidth]{symmetric_N3_Pr0v6.eps}
    \end{minipage}\\
    \vspace{.2cm}
    \centering
    \begin{minipage}[b]{.49\textwidth}
        \centering
        \includegraphics[trim = 0mm 0mm 0mm 0mm,  clip, width=1\columnwidth]{symmetric_N15_Pr0v6.eps}
    \end{minipage}%
    \hfill%
    \begin{minipage}[b]{.49\textwidth}
        \centering
        \includegraphics[trim = 0mm 0mm 0mm 0mm,  clip, width=1\columnwidth]{symmetric_N30_Pr0v2.eps}
    \end{minipage}\\
    \hspace{.005\textwidth}
    \begin{minipage}[t]{0.999\textwidth}
        \caption{Average normalized distortion as a function of the lifetime $n$ with fading. We do not explicitly represent the scenario $N = 15$ and $\overline{\rm Pr}_{\ttx} = 0.2$ because it is analogous to the case $N = 3$ and $\overline{\rm Pr}_{\ttx} = 0.2$. Moreover, the case $N = 30$ and $\overline{\rm Pr}_{\ttx} = 0.6$ has no solutions since too many users are considered and $\overline{\rm Pr}_{\ttx}$ is too high.}
        \label{fig:symmetric}
    \end{minipage}
    \vspace{-.5cm}
\end{figure*}

In this section we show how the system parameters influence the distortion of the network. We consider three groups of nodes placed at different locations from the BS.

The transmission parameters are taken from the datasheets of two real devices, namely
RN-131C 802.11 b/g Wireless LAN Module (a module with extremely low power consumption for Wi-Fi connections) and RC2400HP RF Transceiver Module (an RF module based on ZigBee and IEEE 802.15.4). The former uses a central transmission frequency and bandwidth of $2.441$~GHz and $W = 5$~MHz, respectively. The maximum power used for transmission is $237.7$~mW, and, when a transmission is performed, the minimum and maximum consumed powers are $462$~mW and $699.6$~mW, respectively. When only the RF chain is considered, assuming a minimum transmission power $P_{\min,i} = 100$~mW and according to Section~\ref{subsec:energy_model}, we derive $\eta_{A,i} = 0.58$ and $\mathcal{E}_{C,i} = 167.75$~mW (note that this is an approximation, and the values may slightly change depending on $P_{\min,i}$).

The RC2400HP module uses the same central frequency and bandwidth of the other module, but has different energy related parameters; in particular, $P_{\min,i} = 11.22$~mW, $P_{\max,i} = 107.15$~mW, $\mathcal{E}_{C,i} = 60.15$~mW and $\eta_{A,i} = 0.23$.

If not otherwise stated, we also assume some parameters to be common to all devices, and in particular: the constant energy term $\beta_i$ (see~\eqref{eq:e_circuitry}) is equal to $1$~mJ in every frame; the energy consumption function due to the processing evolves linearly with $L_i^{(k)}$ as in Equation~\eqref{eq:e_processing_lin}, with a slope of $E_{0,i} \,\alpha_{P,i} = 50$~nJ/bit and a coefficient $\beta_{P,i} = 0$~bit$^{-1}$; the channel gains are computed using the standard path loss model with a path-loss exponent equal to $3.5$ (e.g., as in an urban scenario) and are affected by Rayleigh fading; the overall noise power spectral density is $-167$~dBm/Hz; the frame duration $T$ is $1$~s. The parameters of the distortion curves are $a_i = 0.35$ and $b_i = 19.9$, which have been derived empirically fitting the realistic rate-distortion curves of~\cite{zordan}. Moreover, nodes are divided in three groups,  G$_1$, G$_2$ and G$_3$. Nodes in G$_1$ and G$_2$ use the parameters of the RN-131C module, whereas devices in G$_3$ use the RC2400HP module. The groups are heterogeneous in terms of distance from the base station, amount of data to send, and QoS requirements. We maintain these parameters constant during the whole lifetime\footnote{In monitoring IoT applications with periodically sensing devices, it is very likely to have static nodes.}.
The first group includes nodes that are located at a distance $d_{\mathrm{G}_1} = 4$~m from BS (very close to the base station), transmit packets with $L_{0,i}^{(k)} = 2$~Mbit and demand a distortion below $D_{\tth,i}^{(k)} = 8\%$. Nodes in G$_2$ have $d_{\mathrm{G}_2} = 20$~m, $L_{0,i}^{(k)} = 1$~Mbit and looser distortion requirements $D_{\tth,i}^{(k)} = 15\%$. Finally, the third group consists of nodes very far away from BS ($d_{\mathrm{G}_3} = 100$~m), which transmit fewer bits ($L_{0,i}^{(k)} = 10$~kbit), but require better QoS ($D_{\tth,i}^{(k)} = 4\%$). Note that our model is able to handle the heterogeneity of this scenario, which is a key feature of many WSNs.


\emph{\textbf{Distortion vs. lifetime}.} In \figurename~\ref{fig:symmetric}, we plot the distortion and the lifetime obtained as solutions of the optimization problem~\eqref{eq:general_output} for different values of the number of nodes (namely $N \in \{3,15,30\}$), uniformly distributed among the three groups, and a transmission probability $\overline{\rm Pr}_{\ttx} \in \{0.2,0.6\}$. The continuous lines represent the optimal solution described in Section~\ref{sec:FOP_avg_CSI}, which explicitly takes into account the fading effects; instead, the dashed lines are obtained using the approach of Section~\ref{subsec:sub_opt_pol} that assumes the channel gains to be constant during the whole frame, and thus represents a suboptimal policy when the channels are affected by fading. 

The curves have been obtained by changing the weight factor $\sigma$ of Equation~\eqref{eq:general_output}, and the dismission procedure of Section~\ref{subsec:dismission} is not performed here. Also, we consider very high traffic conditions, in which every node has data to transmit at every time frame (this also explains why the lifetime is very short). Although this scenario is not always realistic, it is important to analyze for two reasons: 1) it gives information about the maximum performance the network can achieve, since with lower traffic conditions the distortion cannot get worse, and 2) it can be easily remapped to a more general model with many more nodes and lower traffic patterns. For example, the trend of the curves would be the same in a network with twice the nodes which transmit half the time; however, in this case the network lifetime would be doubled.

The distortion tends to increase with the lifetime, as expected, since smaller amounts of energy can be allocated in each frame and thus nodes must compress more to transmit their data. For small values of $n$, the curves are constant because not all the energy in the batteries is used or because a zero distortion is achieved. Clearly, it is always better to choose the right extremes of the constant regions rather than the other points, as they provide the same QoS with longer lifetimes. Also, it can be noticed that the higher $\overline{\rm Pr}_{\ttx}$, the higher the distortion (i.e., the worse the performance); this happens because higher transmission probabilities impose to transmit more often, and thus to access the channel even when its gain is low (which, in turn, does not allow to transmit many bits). 

\begin{figure}[t]
  \centering
  \includegraphics[trim = 0mm 0mm 0mm 0mm,  clip, width=1\columnwidth]{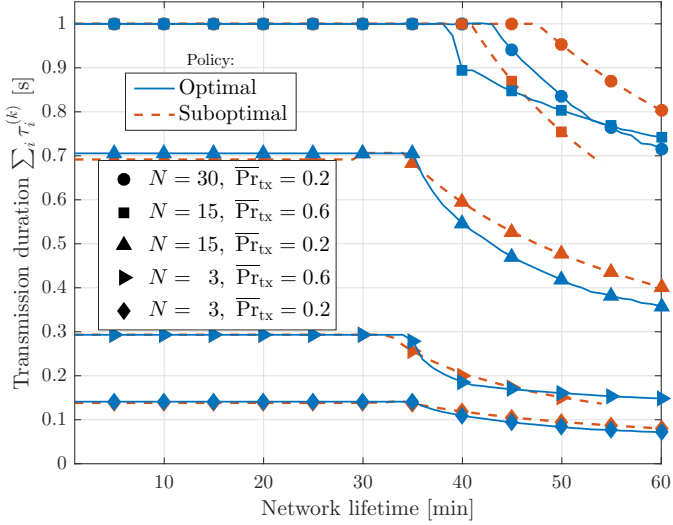}
  \caption{Transmission durations as a function of the lifetime $n$ with fading when $B_{0,i} = 240$~J.}
  \label{fig:tau_symmetric}
\end{figure}

As shown in \figurename~\ref{fig:tau_symmetric}, when $\overline{\rm Pr}_{\ttx}$ is low and $N$ is small, the optimal and suboptimal approaches almost coincide, since a transmission is performed only when the channel gain is very high, and $\sum_i \tau_i^{(k)} \ll T^{(k)}$ (thus it is possible to serve all users even with a suboptimal approach). However, when $\overline{\rm Pr}_{\ttx}$ increases, transmissions are more frequent, whereas when $N$ is larger the sum $\sum_i \tau_i^{(k)}$ may saturate to $T^{(k)}$. In these cases the benefits of adopting the optimal scheme are relevant, and using a suboptimal approach may even lead to an infeasible solution, where instead the optimal scheme is still feasible. 

The maximum lifetime is reached when the problem becomes infeasible (see Section~\ref{subsec:dismission}), i.e., no energy allocation can satisfy all constraints of FOP with an acceptable distortion for all the frames of the considered lifetime. However, note that the curves do not necessarily reach value $1$; this happens because we impose $\overline{\rm Pr}_{\ttx} > 0$ and we guarantee that a feasible distortion is achieved even in the limit condition $\theta_i^{(k)} = \theta_{\ttx,i}^{(k)}$ (see Equation~\eqref{eq:Pr_tx}). Thus, when we take the average over $\theta_i^{(k)} \geq \theta_{\ttx,i}^{(k)}$ (Equation~\eqref{eq:D_avg}), we integrate all terms smaller than one, and therefore we obtain an average distortion $\overline{D}_i^{(k)} < 1$. 
We also remark that, as can be seen in the figure, the lifetime strongly depends on the initial battery level. However, even if larger batteries were considered, the trend of the distortion curves would remain the same.

\emph{\textbf{Transmission duration}.} The overall transmission durations are represented in \figurename~\ref{fig:tau_symmetric}. As previously explained, when $N$ is high (e.g., the case $N = 30$ and $\overline{\rm Pr}_{\ttx} = 0.2$), or if higher transmission probabilities are required (e.g., the case $N = 15$ and $\overline{\rm Pr}_{\ttx} = 0.6$), then $\sum_i \tau_i^{(k)}$ reaches $T^{(k)} = 1$~s. In all the other cases, every node can be considered independently of the others, since they always satisfy the time constraint. Note that $\sum_i \tau_i^{(k)}$ decreases with the lifetime, since less energy is available for transmission at every slot. 

\begin{figure}[t]
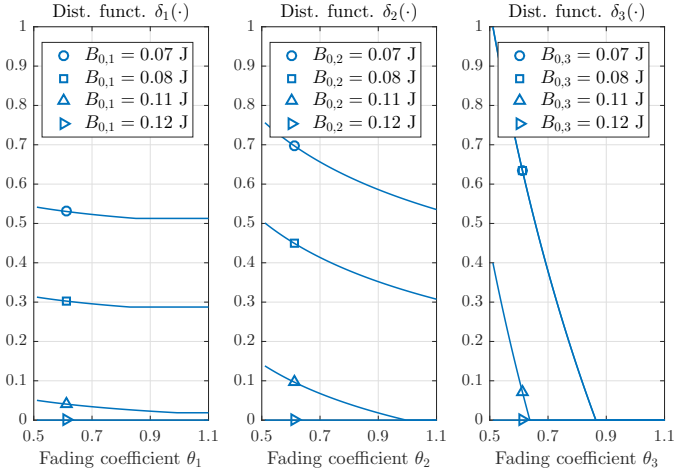

  \centering
  \includegraphics[trim = 0mm 0mm 0mm 0mm,  clip, width=.32\columnwidth]{delta_funct1.eps}~
  \includegraphics[trim = 0mm 0mm 0mm 0mm,  clip, width=.32\columnwidth]{delta_funct2.eps}~
  \includegraphics[trim = 0mm 0mm 0mm 0mm,  clip, width=.32\columnwidth]{delta_funct3.eps}
  \caption{Optimal distortion functions $\delta_i(\cdot)$ as a function of $\theta_i$ for $i \in \mathcal{N} = \{1,2,3\}$ in a single frame when $\overline{\rm Pr}_{\ttx,i} = 0.6$.}
  \label{fig:delta_function}
\end{figure}

\begin{figure}[t]
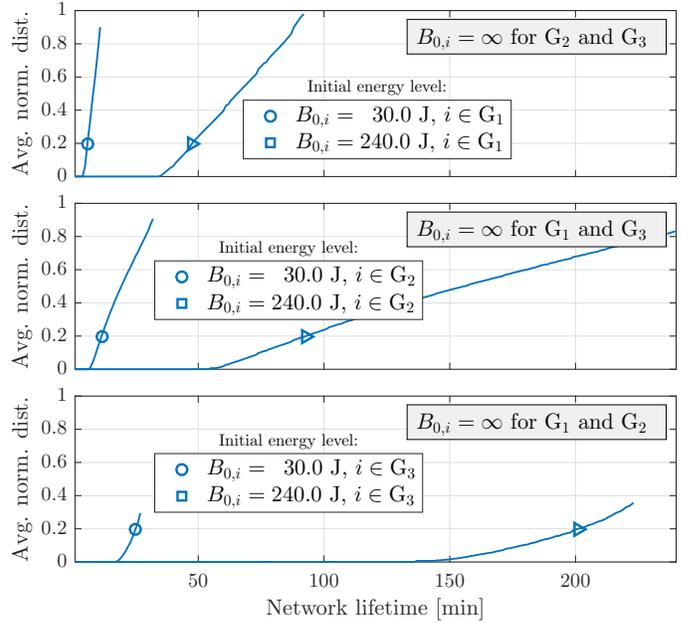

  \centering
  \includegraphics[trim = 0mm 0mm 0mm 0mm,  clip, width=1\columnwidth]{changeL0_group1.eps}\\
  \vspace{.1cm}
  \includegraphics[trim = 0mm 0mm 0mm 0mm,  clip, width=1\columnwidth]{changeL0_group2.eps}\\
  \vspace{.1cm}
  \includegraphics[trim = 0mm 0mm 0mm 0mm,  clip, width=1\columnwidth]{changeL0_group3.eps}
  \caption{Average normalized distortion as a function of the lifetime $n$ with fading when $L_{0,i}^{(k)}$ evolves over time when $N = 3$. In every figure, the initial battery levels of two groups are infinite.}
  \label{fig:changeL0}
\end{figure}

\begin{figure}[t]
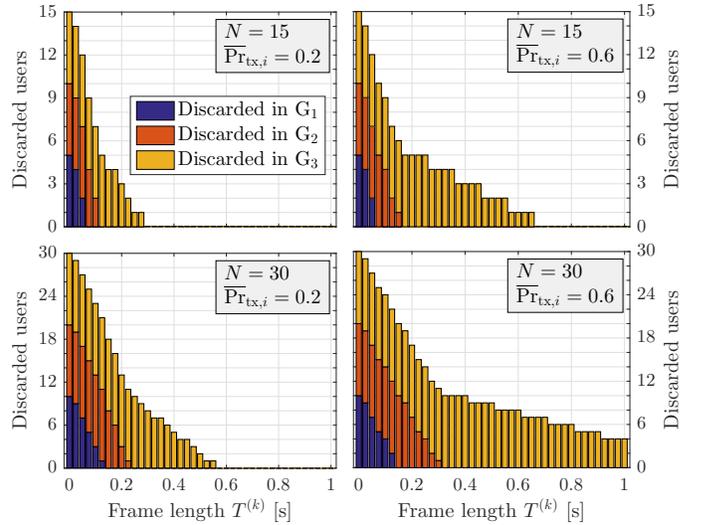

  \centering
  \includegraphics[trim = 0mm 0mm 0mm 0mm,  clip, width=.49\columnwidth]{admit_N15_Pr0v2.eps}~
  \includegraphics[trim = 0mm 0mm 0mm 0mm,  clip, width=.49\columnwidth, height=.345\columnwidth]{admit_N15_Pr0v6.eps}\\
  \vspace{.1cm}
  \includegraphics[trim = 0mm 0mm 0mm 0mm,  clip, width=.49\columnwidth]{admit_N30_Pr0v2.eps}~
  \includegraphics[trim = 0mm 0mm 0mm 0mm,  clip, width=.49\columnwidth, height=.42\columnwidth]{admit_N30_Pr0v6.eps}
  \caption{Number of discarded users as a function of the frame length.}
  \label{fig:admit_map}
\end{figure}

\emph{\textbf{Distortion function}.} In the previous simulations, for every channel realization, we have implicitly computed the transmission power function and the corresponding distortion function $\delta_i(\cdot)$ defined in Equation~\eqref{eq:delta_funct}. Its trend can be seen in \figurename~\ref{fig:delta_function} for the case $N = 3$. As expected, the distortion decreases as $\theta_i$ increases, since this corresponds to better channel conditions. However, the distortion does not reach $0$ in all cases, and may instead saturate to a constant value. This happens because the energy constraints (given by the initial battery levels) are tight, and it is not possible to send long packets even when the channel gain is high. The amount of data sent in the constant regions is $L_{\een,i}^{(k)}$ given by Equation~\eqref{eq:L_en}.

\emph{\textbf{Different packet sizes}.} In \figurename~\ref{fig:changeL0}, the amount of data that the nodes send over time $L_{0,i}^{(k)}$ is modified according to a fixed pattern. Let $L_{0,\mathrm{G}_j}$ be the original packet length previously introduced for group $j = 1,2,3$. Then, in this example we use $L_{0,i}^{(k)} = \zeta^{(k)} L_{0,\mathrm{G}_j}$, with $i \in \mathrm{G}_j$ and a coefficient $\zeta^{(k)}$ that periodically increases and then decreases with $k$; in particular, $\zeta^{(k)}$ evolves as $\frac{1}{2}$, $1$, $2$, $1$, $\frac{1}{2}$, $1$, $2$, etc. (e.g., this models sensors that track a periodic phenomenon). 

In \figurename~\ref{fig:changeL0}a, we assume that the initial energy levels of all nodes in G$_2$ and G$_3$ are infinite, whereas the nodes in G$_1$ have a limited reserve of energy (similarly, in \figurename s~\ref{fig:changeL0}b and \ref{fig:changeL0}c, group G$_2$ or G$_3$ is energy constrained, respectively). It is interesting to note that, in order to guarantee fairness, even if most of the nodes have infinite resources, the network distortion is still limited. Also, the lifetime strongly depends on which group has limited resources: G$_1$ transmits lots of data, hence the system quickly becomes infeasible when its initial energy level is low; instead, nodes in G$_3$ do not transmit large packets, and thus they consume less energy over time and the lifetime is much longer. Note that EAP plays a fundamental role here, since it assigns different amounts of energy according to the packet lengths $L_{0,i}^{(k)}$.

\emph{\textbf{Dismission policy}.} So far, we have neglected the dismission policy described in Section~\ref{subsec:dismission}. However, when the problem is infeasible, it is possible to discard some users and allow the others to transmit. In \figurename~\ref{fig:admit_map} we assume that the priority of the users is given by the group they belong to (G$_1$ $\succ$ G$_2$ $\succ$ G$_3$). As can be seen, when $T^{(k)}$ is larger, fewer users are dismissed, because satisfying Constraint~\eqref{eq:FOP_tau} is easier. Also, note that in the case $N = 30$ and $\overline{\rm Pr}_{\ttx} = 0.6$, discarding some users is necessary even if $T^{(k)} = 1$~s. This explains why in \figurename~\ref{fig:symmetric} we could not represent this setup, since the dismission policy was not applied.

\begin{figure}[t]
  \centering
  \includegraphics[trim = 0mm 0mm 0mm 0mm,  clip, width=1\columnwidth]{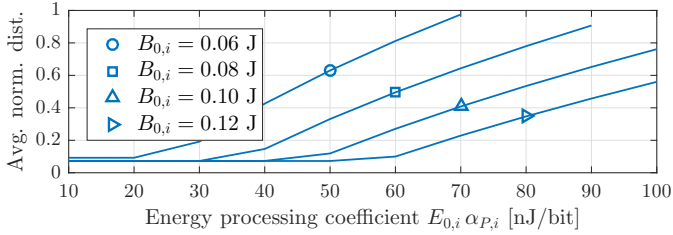}
  \caption{Average normalized distortion as a function of the the energy processing coefficient of Equation~\eqref{eq:e_processing_lin} when $N = 3$.}
  \label{fig:E_params}
\end{figure}

\emph{\textbf{Processing energy}.} In our numerical results, we adopt Equation~\eqref{eq:e_processing_lin} to model the processing energy. In practice, this parameter has a strong impact on the system distortion, as can be observed in \figurename~\ref{fig:E_params}. This is mainly due to groups G$_1$ and G$_2$, which transmit long packets and thus consume much more energy.

\section{Conclusions}\label{sec:conclusions} 

In this work, we presented a dynamic TDMA-based scheduling that permits the joint optimization of energy consumption and data distortion. We studied the tradeoff between lifetime and distortion, and set up a framework that optimizes the energy to allocate in every slot, determines the compression degree of the data to send along with the transmission durations, and performs power control.
We  accounted for two different levels of channel knowledge, i.e., when the fast fading realization 
is or is not considered, resulting in a more or less robust allocation policy, respectively. The thorough numerical evaluation based on the characteristics of realistic devices validates the analytical results and shows that using the optimal approach with dynamic power control outperforms simpler schemes.

Future work includes the extension of the model to processing energy consumption functions that decrease with the compression ratio (as described in Section~\ref{subsec:energy_model}), the study of latency requirements on the data reception, and the analysis of packet losses on the network performance, since they may have a strong impact on both the energy and the distortion metrics.

\section*{Acknowledgments}
This work was partially supported by Intel's Corporate Research Council.

\appendices

\section{Proof of Lemma~\ref{lemma:FOP_gamma}}\label{proof:FOP_gamma}
By contradiction, assume that there exists a user $\ell \in \mathcal{N}$ for which a distortion $D_\ell^\prime < \Gamma^\star D_{\tth,\ell}$ leads to an optimal solution, but $D_\ell^{\prime\prime} = \Gamma^\star D_{\tth,\ell}$ does not. According to~\eqref{eq:dist}, this implies that longer (higher quality) packets are used in the first case (i.e., $L_\ell^\prime > L_\ell^{\prime \prime}$). Then, if we chose $P_{\ttx,\ell}^{\prime\prime} = P_{\ttx,\ell}^\prime$ and $\tau_\ell^{\prime\prime} = \tau_\ell^\prime L_\ell^{\prime\prime} / L_\ell^\prime < \tau_\ell^\prime$, all constraints~\eqref{eq:FOP_distortion}-\eqref{eq:FOP_tau} would still be satisfied (note that with longer packets the energy consumption is higher) and we would find a feasible optimal solution in which~\eqref{eq:FOP_gamma} is satisfied with equality, i.e., the initial assumption must be wrong.

\section{Proof of Theorem~\ref{thm:Gamma1_Gamma2}}\label{proof:Gamma1_Gamma2}
Consider a generic node $\ell$. According to the procedure described in Section~\ref{subsec:FOP_full_CSI} to solve Problem~\eqref{prob:g}, the point $P_{\ttx,\ell}^\star$ (if defined) either falls in the increasing right branch of $g_\ell(x)$ or is equal to $P_{{\rm max},\ell}$.
        
By assumption we have $\Gamma^\prime \leq \Gamma^{\prime\prime}$, which, using Definition~\eqref{eq:dist}, implies $L_\ell^\prime \geq L_\ell^{\prime\prime}$. Since we focus on compression algorithms with an energy consumption proportional to the compressed packet size (see Section~\ref{subsec:energy_model}), the RHS of~\eqref{eq:g_Pi} is a decreasing function of $L_\ell$. Thus, by naming $P^\prime$ and $P^{\prime\prime}$ the points $\overline{P}_\ell$ corresponding to $\Gamma^\prime$ and $\Gamma^{\prime\prime}$, respectively, we obtain that 1) $P^{\prime\prime}$ exists, since $P^\prime \geq P_{{\rm min},\ell}$ exists by assumption, and 2) $P^\prime \leq P^{\prime\prime}$. 
        
Finally, since the same energy $E_\ell$ is considered in the two cases, $P^\prime$ leads to a transmission duration longer than it would be with $P^{\prime\prime}$, i.e., $\tau_\ell^{\prime} \geq \tau_\ell^{\prime\prime}$, and therefore Constraint~\eqref{eq:FOP_tau} remains satisfied also with $\Gamma^{\prime\prime}$. 
In conclusion, since all the constraints of Problem~\eqref{prob:FOP} were satisfied with $P^\prime$, they still hold also when using $P^{\prime\prime}$.

\section{Proof of Lemma~\ref{lemma:delta_tau}}\label{proof:delta_tau}
We equivalently prove that $L_i^{\theta_i} \triangleq  \tau_iW \log_2(1+h_{0,i} \,\theta_i\, \rho_i(\theta_i))$ first increases and then decreases in $\tau_i$ (see~\eqref{eq:dist}). Note that $\rho_i(\theta_i)$ intrinsically depends on $\tau_i$, according to Problem~\eqref{prob:g_theta}.
        
First, consider the case $P_{\min,i} \leq \smash{\overline{P}}_i^{\theta_i} \leq P_{\max,i}$ in~\eqref{prob:g_theta}, and thus $\rho_i(\theta_i) = \smash{\overline{P}}_i^{\theta_i}$ for $\tau_i$ and~\eqref{eq:fading_EC_less_Ei} is satisfied with equality. Then, we can rewrite $\tau_i$ and $L_i^{\theta_i}$ as~
\begin{align}
    \tau_i & = \frac{E_i-\beta_i}{E_{P,i}(\rho_i(\theta_i)) + \dfrac{\rho_i(\theta_i)}{\eta_{A,i}} + \mathcal{E}_{C,i}} \label{eq:proof_tau_i_fading} \\
    \begin{split} \label{eq:proof_L_g_theta}
        L_i^{\theta_i} & = \frac{W \, (E_i-\beta_i)\,\log_2(1+h_{0,i}\,\theta_i\,\rho_i(\theta_i))}{E_{P,i}(\rho_i(\theta_i)) + \dfrac{\rho_i(\theta_i)}{\eta_{A,i}} + \mathcal{E}_{C,i}} \\
        & = \frac{W (E_i-\beta_i)}{k_i^{\theta_i}  + g_i^{\theta_i}(\rho_i(\theta_i))}
    \end{split}
\end{align}

\noindent where $k_i^{\theta_i} \triangleq E_{P,i}(\rho_i(\theta_i))/\log_2(1+h_{0,i}\,\theta_i\,\rho_i(\theta_i)) $ is a constant term, and function $g_i^{\theta_i}(\cdot)$ is analogous to $g_i(\cdot)$ defined in~\eqref{eq:g_Pi} when the fading gain is $\theta_i$. By exploiting the structure of $g_i^{\theta_i}(\cdot)$ (see \figurename~\ref{fig:plot_g}) and considering that the other terms of~\eqref{eq:proof_L_g_theta} are fixed, we have that $L_i^{\theta_i}$ initially increases and then decreases with $\rho_i(\theta_i)$, for a given $\tau_i$. Moreover, since $\tau_i$ is monotonic in $\rho_i(\theta_i)$ (Equation~\eqref{eq:proof_tau_i_fading}), also $\rho_i(\theta_i)$ is monotonic in $\tau_i$, and we have that $L_i^{\theta_i}$ initially increases and then decreases in $\tau_i$. 

If $\smash{\overline{P}}_i^{\theta_i} > P_{\max,i}$, then $\rho_i(\theta_i) \equiv P_{\max,i}$ and~\eqref{eq:fading_EC_less_Ei} is satisfied with strict inequality. 
Since $\rho_i(\theta_i)$ is constant, $L_i^{\theta_i}$ is monotonically increasing in $\tau_i$. As $\tau_i$ increases, also the RHS of~\eqref{eq:proof_tau_i_fading} increases, until the constraint is satisfied with equality and thus we fall again in the previous case, where $\smash{\overline{P}}_i^{\theta_i} <P_{\max,i} $ and $L_i^{\theta_i}$ initially increases and then decreases with $\tau_i$.

Finally, if $ \smash{\overline{P}}_i^{\theta_i} < P_{\min,i}$, either $\rho_i(\theta_i)$ is not defined for the considered $\tau_i$ and this case is not of interest, or $\rho_i(\theta_i) \equiv P_{\min,i}$ and $L_{\een, i}$ bits are transmitted. In the latter case, $L_i^{\theta_i} \equiv L_{\een,i}$, and~\eqref{eq:fading_EC_less_Ei} is satisfied with equality, thus $L_i^{\theta_i}$ decreases as $\tau_i$ increases. 
As $\tau_i$ increases, the term $\beta_i + ({P_{\min,i}}/{\eta_{A,i}} + \mathcal{E}_{C,i})\, \tau_i$ will exceed $E_i$, making the problem infeasible. 
As $\tau_i$ decreases, $L_i^{\theta_i}$ increases, until the point in which $\smash{\overline{P}}_i^{\theta} \geq P_{\min,i}$ can be satisfied, and thus we go back to the previous case, where $L_i^{\theta_i}$ is initially increasing and then decreasing in $\tau_i$.

\section{Proof of Theorem~\ref{thm:alternate_opt}}\label{proof:alternate_opt}

The energy allocation optimization problem is convex, but non-strictly convex in general. This means that a unique minimum, namely $\Psi^\star$, exists but, potentially, with multiple minimum points.         
We now show that Algorithm~\ref{alg:AO} produces a non-increasing sequence of distortion values ($\Psi_1 \geq \Psi_2 \geq\ldots$), and, unless the minimum is achieved, there exists a non-zero probability that $\Psi_m > \Psi_{m+1}$, for some $m$. Since the minimum is unique, the sequence converges to $\Psi^\star$ thanks to~\cite[Proposition~2.7.1]{bertsekas}.

        
Consider the generic step $m$ of the algorithm, in which the optimization revolves around node $\ell$, i.e., we focus on the sequence $\mathbf{E}_\ell$. As described in Section~\ref{subsec:FOP_convex}, the objective $f_{\rm FOP}^{(k)}$ is a convex function of the optimization variable $E_\ell^{(k)}$. In particular, $f_{\rm FOP}^{(k)}$ is strictly convex in $[\underline{E}_\ell^{(k)},\smash{\overline{E}}_\ell^{(k)}]$, and constant for $E_\ell^{(k)} \geq \smash{\overline{E}}_\ell^{(k)}$. 

Since the sum of convex functions is convex, when we aim at optimizing $\sum_{k = 1}^n f_{\rm FOP}^{(k)}$ in $[\underline{\mathbf{E}}_\ell,\overline{\mathbf{E}}_\ell]$, this is a convex optimization problem and can be solved using~\eqref{prob:EAP_red_L}. Let $\mathbf{E}_\ell^\star$ be its solution. The following two cases can occur.
\begin{enumerate}
    \item All the elements of $\mathbf{E}_\ell^\star$ fall inside the strictly convex region $[\underline{\mathbf{E}}_\ell,\overline{\mathbf{E}}_\ell)$ ($\overline{\mathbf{E}}_\ell$ is excluded), e.g., because of the battery constraints. In this case, there is only one optimal solution, and no other actions are required for the current step ($S = 0$ in Line~\ref{alg:line:S} of the algorithm);
    \item Part of the elements of $\mathbf{E}_\ell^\star$ fall at the beginning of the constant region, i.e., $E_\ell^{(k)} = \smash{\overline{E}}_\ell^{(k)}$ for some $k \in \overline{\mathcal{K}}$. In this case, also other solutions may be optimal for the current iteration, since all the feasible energy combinations with $E_\ell^{(k)} \succeq \smash{\overline{E}}_\ell^{(k)}$ for $k \in \overline{\mathcal{K}}$ lead to the same solution. However, although all these sequences provide the same $\Psi_m$ at the current step, they may influence the future values $\Psi_{m+1},\Psi_{m+2},$ etc. Then, two subcases should be distinguished: in the following $N-1$ steps of the algorithm (i.e., the next time that node $\ell$ is examined), the sequence of $\Psi$ has either strictly decreased, or remained constant. In the former case, the algorithm proceeds toward the optimal solution. In the latter, the algorithm cyclically returns to point $m$, thus the sequence of $\Psi$ has not improved. In this case, we choose other points $E_\ell^{(k)} > \smash{\overline{E}}_\ell^{(k)},\, k \in \overline{\mathcal{K}}$ (e.g., with a random approach as described in Lines~\ref{alg:line:v_vect}-\ref{alg:line:end_AO}) and repeat the procedure. In an infinite horizon, all the possible energy combinations have been tested with a non-null probability, thus the algorithm has proceeded toward an optimal solution.
\end{enumerate}

In the worst case scenario, Algorithm~\ref{alg:AO} may degenerate in almost an exhaustive search; however, we have observed that, in practical cases, very few iterations are required, and the algorithm rapidly converges.

\bibliographystyle{IEEEtran}
\bibliography{bib}

\end{document}